%% file: template.tex
\documentclass{article}

\usepackage{arxiv}

\usepackage[utf8]{inputenc} 
\usepackage[T1]{fontenc}    
\usepackage{hyperref}       
\usepackage{url}            
\usepackage{booktabs}       
\usepackage{amsfonts}       
\usepackage{nicefrac}       
\usepackage{microtype}      
\usepackage{lipsum}		
\usepackage{graphicx}
\usepackage{doi}
\usepackage{makecell}
\usepackage{multirow}
\usepackage{colortbl}
\usepackage{hhline}
\usepackage{rotating}
\usepackage{tabularray}
\usepackage{enumitem}
\usepackage{amsmath}
\usepackage{cleveref}
\usepackage{authblk}
\usepackage{subcaption}
\usepackage[table]{xcolor}   
\usepackage{tabularx}        
\usepackage{array}           

\definecolor{softblue}{HTML}{e3f5ff}
\definecolor{my_grey}{rgb}{0.945,0.945,0.945}
\definecolor{my_blue}{rgb}{0.898,0.929,1}
\definecolor{bibgray}{rgb}{0.3, 0.3, 0.3}

\newcolumntype{C}[1]{>{\centering\arraybackslash}m{#1}}
\newcolumntype{Y}{>{\centering\arraybackslash}X}

\title{Application of Tabular Transformer Architectures for Operating System Fingerprinting}

\date{} 					

\author{
  \textbf{Rubén Pérez-Jove}\textsuperscript{1,2,*}\href{https://orcid.org/0000-0002-7988-945X}{\includegraphics[scale=0.06]{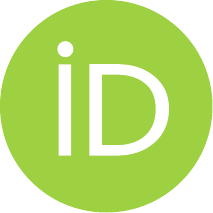}} , 
  \textbf{Cristian R. Munteanu}\textsuperscript{1,2,3}\href{https://orcid.org/0000-0002-5628-2268}{\includegraphics[scale=0.06]{orcid.pdf}} , 
  \textbf{Alejandro Pazos}\textsuperscript{1,2,3}\href{https://orcid.org/0000-0003-2324-238X}{\includegraphics[scale=0.06]{orcid.pdf}} , 
  \textbf{Jose Vázquez-Naya}\textsuperscript{1,2}\href{https://orcid.org/0000-0002-6194-5329}{\includegraphics[scale=0.06]{orcid.pdf}} \\
  \textsuperscript{1} RNASA-IMEDIR Research Group, Department of Computer Science and Information Technologies, \\ 
  Facultad de Informática, Universidade da Coruña, Elviña, 15071 A Coruña, Spain \\ 
  \texttt{\{c.munteanu,alejandro.pazos,jose\}@udc.es} \\ \vspace{5pt}
  \textsuperscript{2} CITIC Research Centre, Universidade da Coruña, Elviña, 15071 A Coruña, Spain \\ \vspace{5pt}
  \textsuperscript{3} IKERDATA S.L, ZITEK, University of Basque Country UPVEHU, Rectorate Building, 48940 Leioa, Spain \\ \vspace{5pt}
  * \textit{Corresponding author}: \texttt{ruben.perez.jove@udc.es}
}

\renewcommand{\shorttitle}{Application of Tabular Transformer Architectures for Operating System Fingerprinting}

\hypersetup{
pdftitle={Application of Tabular Transformer Architectures for Operating System Fingerprinting},
pdfsubject={cs.AI, cs.CR, cs.NI},
pdfauthor={Rubén Pérez-Jove; Alejandro Pazos; Jose Vázquez-Naya},
pdfkeywords={Operating System; Fingerprinting; Identification; Detection; Deep Learning; Transformer; FT-Transformer; TabTransformer; Machine Learning; Cybersecurity},
}

\begin{document}
\maketitle


\begin{abstract}

Operating System (OS) fingerprinting is essential for network management and cybersecurity, enabling accurate device identification based on network traffic analysis. Traditional rule-based tools such as Nmap and p0f face challenges in dynamic environments due to frequent OS updates and obfuscation techniques. While Machine Learning (ML) approaches have been explored, Deep Learning (DL) models, particularly Transformer architectures, remain unexploited in this domain. This study investigates the application of Tabular Transformer architectures—specifically TabTransformer and FT-Transformer—for OS fingerprinting, leveraging structured network data from three publicly available datasets. Our experiments demonstrate that FT-Transformer generally outperforms traditional ML models, previous approaches and TabTransformer across multiple classification levels (OS family, major, and minor versions). The results establish a strong foundation for DL-based OS fingerprinting, improving accuracy and adaptability in complex network environments. Furthermore, we ensure the reproducibility of our research by providing an open-source implementation.

\end{abstract}




\input{content/s01-introduction}
\input{content/s02-background}
\input{content/s03-related-work}
\input{content/s04-materials-methods}
\input{content/s05-results-discussion}
\input{content/s06-conclusion}


\section*{Acknowledgments}
\label{sec:ack}

This work was supported by the grant \textbf{ED431C 2022/46 – Competitive Reference Groups GRC} – funded by: EU and Xunta de Galicia (Spain). This work was also supported by \textbf{CITIC}, funded by Xunta de Galicia through the collaboration agreement between the Consellería de Cultura, Educación, Formación Profesional e Universidades and the Galician universities to strengthen the research centres of the Sistema Universitario de Galicia (CIGUS); and by the \textbf{``Formación de Profesorado Universitario'' (FPU)} grant from the Spanish Ministry of Universities to \textbf{Rubén Pérez Jove (Grant FPU22/04418)}. This work was also made possible through the access granted by the \textbf{Galician Supercomputing Center (CESGA)} to its supercomputing infrastructure. The supercomputer FinisTerrae III and its permanent data storage system have been funded by the Spanish Ministry of Science and Innovation, the Galician Government and the European Regional Development Fund (ERDF).


\bibliographystyle{IEEEtran}
\bibliography{references} 

\end{document}

%% file: content/s01-introduction.tex
\section{Introduction}
\label{sec:intro}

The ability to accurately identify the characteristics of a host through the analysis of its network traffic is crucial for a variety of tasks in network management and computer security. Accurately identifying a machine’s Operating System (OS) family and version is critical for applications including vulnerability exploitation, network inventory, and the detection of unauthorized devices.

The process of OS fingerprinting entails determining information related to the operating system—such as OS family and version—of a network-connected device by analysing its traffic. These techniques leverage differences arising from the unique ways each OS implements the communication protocol stack. Depending on the approach, OS fingerprinting can be conducted through active or passive scanning. Active scanning involves sending probes to the target and analysing responses, providing speed and reliability at the cost of a higher risk of detection. A widely used tool for this method is Nmap \cite{nmaporg_nmap_nodate}. On the other hand, passive scanning examines existing network traffic without direct interaction with the target, making it a stealthier, though generally slower and less effective, method. Tools such as p0f \cite{zalewski_p0f_nodate} are commonly employed for passive OS detection.

Traditional rule-based approaches, as used by the aforementioned tools, are highly sensitive to variations in machine characteristics. In today’s environment—characterised by a proliferation of connected devices, diverse OSs, and frequent updates—these variations pose significant challenges. An optimal solution would accurately infer a machine's OS even in scenarios where it is newly released, recently updated, or reconfigured. Artificial Intelligence (AI) techniques have demonstrated significant potential in addressing these challenges. Numerous studies have explored the application of AI to OS fingerprinting in recent years \cite{lastovicka_passive_2023}, though most employ classical methods such as typical Machine Learning (ML) algorithms. Research on more advanced techniques, specifically Deep Learning (DL) architectures, remains limited.

\paragraph{Research Problem.} Current OS fingerprinting methods, largely based on traditional rule-based or classical ML approaches, struggle to adapt to the heterogeneity and dynamism of modern network environments. This study seeks to address the problem of how to improve OS fingerprinting accuracy and robustness by leveraging advanced DL architectures—specifically, Transformer-based models designed for tabular data. By doing so, we aim to overcome the limitations of existing methods and provide a solution that is more resilient to evolving network conditions and modern OS variations.

Among recent DL architectures, the Transformer, introduced by Vaswani et al. in 2017 \cite{vaswani_attention_2023}, stands out for its ability to process sequential data efficiently through parallel processing and self-attention mechanisms. This architecture has revolutionised Natural Language Processing (NLP) with the emergence of Large Language Models (LLMs) and has been successfully adapted to other domains, including computer vision with the Vision Transformer (ViT) \cite{dosovitskiy_image_2021}. Its scalability and capacity for generalisation suggest that applying Transformer-based models to OS fingerprinting could yield similar breakthroughs.

A key advantage of the application of Transformers to network traffic data is their ability to capture complex interdependencies. Unlike traditional ML methods that focus on isolated features or require manual importance measures, Transformers use self-attention to process the entire data structure at once. This enables them to dynamically weigh each feature’s contribution, capturing nuanced interactions essential for characterising heterogeneous, dynamic traffic, and ultimately leads to improved OS fingerprinting performance.

\paragraph{Contribution.} In this paper, we propose the application of the Transformer architecture to OS fingerprinting. Given that network traffic data is typically stored as network flows (which can be processed as tabular data), we specifically employ a variant designed for this format: the Tabular Transformer. We analyse two variations of this architecture, namely the TabTransformer (TabT) and the FT-Transformer (FT-T), both optimised for structured tabular data processing.

To rigorously evaluate our approach, we apply it to three publicly available datasets with distinct characteristics that enable classification at multiple granular levels (OS family, major, and minor versions) under different network conditions and feature distributions. We benchmark our models against three representative ML algorithms—k-Nearest Neighbours (kNN), Random Forest (RF), and Multi-Layer Perceptron (MLP)—and compare our results with previous AI-based studies. 

This study makes three key contributions:  
\begin{itemize}  
    \item \textbf{First application of the Transformer architecture to OS fingerprinting:} We introduce attention-based DL models for OS identification, marking the first attempt to apply Transformers—via their adaptation in Tabular Transformers—to this domain.  
    \item \textbf{Comprehensive evaluation across multiple datasets and classification granularities:} We rigorously assess our approach using three publicly available datasets with diverse characteristics, evaluating OS classification at different levels (family, major, and minor versions) and benchmarking against classical ML models and prior research.  
    \item \textbf{Reproducibility and transparency of experiments:} We promote further research by publicly releasing our experimental code under a GNU GPL v3.0 license. This includes preprocessing steps, model implementations, and evaluation metrics, available at: \url{https://github.com/rubenpjove/tabularT-OS-fingerprinting}.  
\end{itemize}  

This paper is structured as follows: \Cref{sec:background} provides essential context on OS fingerprinting, ML models and the Transformer architecture; \Cref{sec:related-work} reviews previous works in both traditional and AI-based OS fingerprinting, including the use of Transformer to other network-related tasks; \Cref{sec:materials-methods} details the experimental design and datasets used; \Cref{sec:results-discussion} presents the results and comparisons with existing approaches; and \Cref{sec:conclusion_future_work} summarises the findings and outlines future research directions.

\clearpage

%% file: content/s02-background.tex
\section{Background}
\label{sec:background}

\subsection{Operating System Fingerprinting}
\label{subsec:os-fingerprinting}

\begin{table}[b!]
    \centering
    \caption{Examples of Window Size and TTL values for different OSs. Source: \cite{lastovicka_jungle_2018}}
    \vspace{5pt}
        \begin{tabular}{>{\centering\arraybackslash}m{2cm}>{\centering\arraybackslash}m{1,5cm}>{\centering\arraybackslash}m{2,5cm}}
            \hline
            \textbf{Window Size} & \textbf{TTL} & \textbf{OS} \\ \hline
            {\cellcolor{my_grey}}8,192 & {\cellcolor{my_grey}}128 & {\cellcolor{my_grey}}Windows 10 \\
            65,535 & 64 & Android 6 \\
            {\cellcolor{my_grey}}29,200 & {\cellcolor{my_grey}}64 & {\cellcolor{my_grey}}Ubuntu \\
            65,535 & 64 & Mac OS X 10.12 \\
            {\cellcolor{my_grey}}65,535 & {\cellcolor{my_grey}}64 & {\cellcolor{my_grey}}iOS 10.3 \\ \hline
        \end{tabular}
    \label{tab:ttls}
\end{table}

As previously introduced, Operating System (OS) fingerprinting is the process of identifying the OS running on a network-connected device by analysing its network traffic characteristics. This technique relies on the fact that different OS implementations exhibit distinct behaviours in network communication, such as variations in Transmission Control Protocol/Internet Protocol (TCP/IP) stack parameters, packet structure, or protocol handling. By examining these characteristics, OS fingerprinting enables the classification of devices, providing valuable information such as OS family, version, or even specific configurations.

As we have already seen, OS fingerprinting plays a crucial role in both network management and security \cite{nmap_os_detection}. It allows administrators to maintain an up-to-date inventory of devices, identify and patch vulnerable systems, and detect unauthorized devices such as rogue access points or insecure personal devices. In cybersecurity, it is widely used for reconnaissance, as identifying a target’s OS enables security professionals to tailor exploits to specific vulnerabilities, increasing the likelihood of a successful attack. Moreover, adversaries can leverage this information for social engineering, impersonating technical support and manipulating users into installing malicious software.

A specific example of OS fingerprinting can be achieved by analysing the Time To Live (TTL) and TCP Window Size parameters in network packets. Different OSs use distinct default values for these parameters, allowing generating specific signatures based on packet observations. Table~\ref{tab:ttls} provides the relation of Window Size and TTL values for various OSs, which enables the classification process. For instance, if a network packet is observed with a TTL of 128 and a TCP Window Size of 8,192, it is likely originating from a Windows 10 machine. These characteristics, combined with other parameters such as TCP options, provide valuable insights for classifying OS versions more accurately.

The level of detail that can be extracted through OS fingerprinting depends on the techniques used and the amount of network data available for analysis. When only basic network parameters, such as TTL or TCP Window Size, are available, OS identification is typically limited to broad categories, and similar OSs can be mixed, as shown in Table~\ref{tab:ttls}. However, incorporating more detailed protocol features allows for finer-grained classification. Depending on the level of detail in the classification, OS fingerprinting can be structured into the following levels:

\begin{itemize}
    \item \textbf{Manufacturer:} Identifies the company or organization responsible for developing the OS, such as Microsoft or Apple.
    \item \textbf{Family:} Classifies the OS into a broader category or series, such as Windows, macOS, or Linux, grouping similar systems under a common architecture.
    \item \textbf{Major Version:} Specifies the primary release version within an OS family, such as Windows 10 or macOS Catalina, indicating significant changes in functionality and system architecture.
    \item \textbf{Minor Version:} Differentiates smaller updates, builds, or patches within a major version, such as Windows 10 version 1909 or macOS Catalina 10.15.5, allowing for more granular classification.
\end{itemize}

\begin{figure}[t!]
    \centering
    \includegraphics[width=\linewidth]{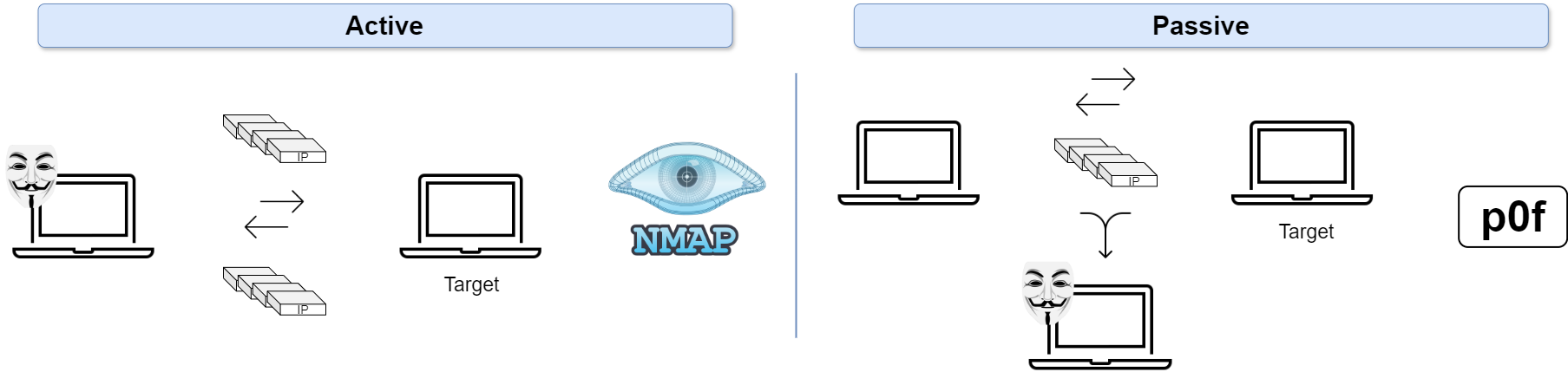}    
    \caption{Diagram of active and passive OS fingerprinting}
    \label{fig:active-passive}
\end{figure}

Based on how the scanner interacts with the network, OS fingerprinting methods are classified into two main categories: active and passive. Active fingerprinting involves sending crafted network packets to the target machine and analysing its responses. While it provides fast and accurate results, it is easily detectable and can be blocked by security mechanisms. A widely well-known tool for active fingerprinting is Nmap \cite{nmaporg_nmap_nodate}, but there are other examples such as Xprobe2 and SinFP \cite{nmap_os_detection, xprobe2, sinfp}. In contrast, passive fingerprinting analyses existing network traffic without directly interacting with the target, making it stealthier but generally less accurate. This method relies on observing characteristics such as TTL, TCP Window Size, and TCP Options. Examples of passive fingerprinting tools include p0f, PRADS, and Ettercap \cite{zalewski_p0f_nodate, eb_fjellskal_prads_2009, ornaghi_alberto_ettercap_2001}. A diagram of both types of OS fingerprinting is shown in Figure~\ref{fig:active-passive}. Furthermore, a detailed analysis of traditional OS fingerprinting methods and tools is presented in Subsection~\ref{subsubsec:traditional-os-fingerprinting}. 

While effective in many scenarios, traditional rule-based OS fingerprinting approaches face significant limitations. Frequent OS updates and patches can alter network signatures, reducing the accuracy of traditional fingerprinting techniques. Additionally, many modern systems implement security mechanisms, such as firewall rules, packet obfuscation, and OS hardening tools, which modify network responses to evade detection. These factors make it increasingly difficult to maintain up-to-date and comprehensive fingerprinting databases, limiting their ability to correctly identify newer OS versions or systems with non-standard configurations.

To overcome these challenges, AI-driven techniques, including ML and DL, have been introduced to enhance OS fingerprinting. By integrating AI techniques, modern OS fingerprinting systems improve the accuracy, handle encrypted or obfuscated traffic, and generalize better across different network conditions. The application of ML and DL to OS fingerprinting is explored in detail in Subsections \ref{subsubsec:ml-based-os-fingerprinting} and \ref{subsubsec:dl-based-os-fingerprinting}.

\subsection{Artificial Intelligence}
\label{subsec:AI}

As background for this work, we provide an overview of the AI methods that form the foundation of our study. Initially, we introduce several conventional ML algorithms that serve as baselines for performance comparison. We then detail the Transformer architecture along with its specialized adaptation for structured tabular data—Tabular Transformers—which constitutes the core of our novel contribution.

\subsubsection{Machine Learning Baselines}
\label{subsubsec:ML-baselines}

For baseline comparison, we used three established algorithms—k-Nearest Neighbors (kNN), Random Forest (RF), and Multi-layer Perceptron (MLP)—which are representative of classical ML paradigms and provide a robust benchmark for evaluating performance on this task.

\begin{itemize}
    \item \textbf{k-Nearest Neighbors (kNN)} \cite{cover_nearest_1967}: A non-parametric method that assigns class labels based on the majority vote among the $k$ closest training examples using distance metrics (e.g., Euclidean, Manhattan). Its simplicity is counterbalanced by increased computational cost on large datasets.
    
    \item \textbf{Random Forest (RF)} \cite{breiman_random_2001}: An ensemble technique that builds multiple decision trees on bootstrapped subsets with random feature selection. The final prediction is obtained by majority voting (classification) or averaging (regression), enhancing generalization and mitigating overfitting.
    
    \item \textbf{Multi-layer Perceptron (MLP)} \cite{rumelhart_learning_1986}: A feed-forward neural network comprising an input layer, one or more hidden layers, and an output layer. Trained via backpropagation, the MLP learns complex nonlinear relationships but requires careful hyperparameter tuning and entails higher computational cost.
\end{itemize}

\subsubsection{The Transformer Architecture}
\label{subsubsec:transformers}

\begin{figure}[t!]
    \centering
    \includegraphics[width=0.4\linewidth]{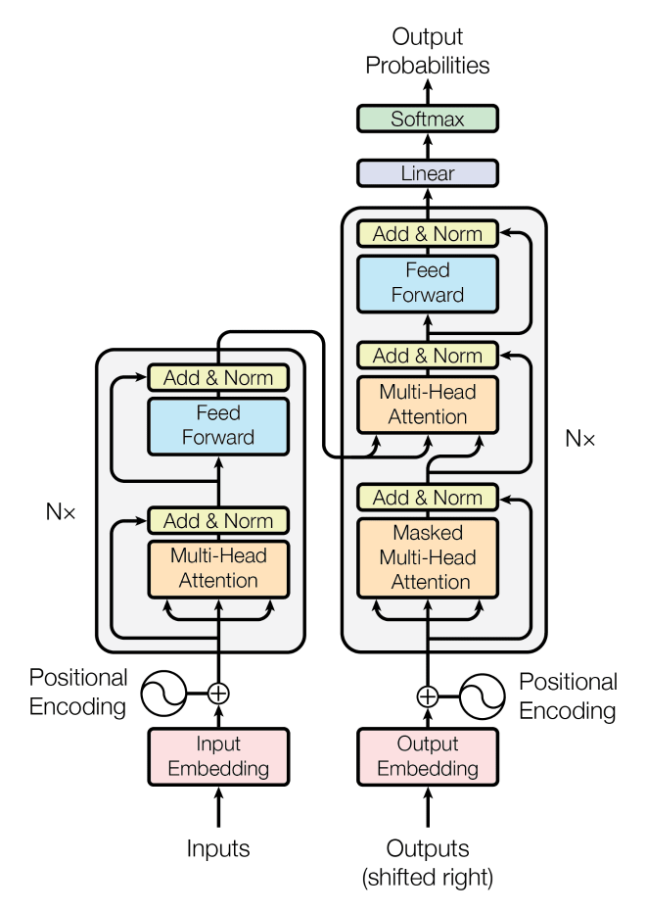}    
    \caption{Transformer architecture. Source: \cite{vaswani_attention_2023}}
    \label{fig:transformer}
\end{figure}

Transformers \cite{vaswani_attention_2023} are a DL architecture specifically designed to process sequential data efficiently. Unlike traditional architectures such as Recurrent Neural Networks (RNNs) and Convolutional Neural Networks (CNNs), Transformers eliminate the need for recurrence and convolutional layers, instead leveraging an encoder-decoder structure, as depicted in \Cref{fig:transformer}. At the heart of this architecture lies the multi-head self-attention mechanism, which allows the model to capture global dependencies across input sequences by dynamically assigning different importance weights to each token. Unlike traditional ML algorithms that rely on extensive feature engineering and domain-specific tuning, Transformers automatically learn complex relationships and interactions within the data. This capability to model long-range dependencies makes them particularly effective for sequence-to-sequence tasks.

Since their introduction in 2017 by Google researchers \cite{vaswani_attention_2023}, Transformers have revolutionized NLP, excelling in tasks such as machine translation, text summarization, and sentiment analysis. A key advantage of this architecture over RNNs is its parallel processing capability, allowing entire input sequences to be processed simultaneously rather than sequentially. This significantly reduces training time and enhances performance, particularly in capturing long-range dependencies within textual data. Consequently, Transformers have become the foundation of modern LLMs, such as OpenAI's GPT and Meta's LLaMA series, which leverage this architecture to achieve state-of-the-art performance in NLP tasks.

Beyond NLP, the adaptability of Transformers has led to their application in diverse domains, including Computer Vision (CV), Reinforcement Learning (RL), or network traffic analysis. Their ability to model complex relationships in structured data makes them suitable for various tasks, from protein structure prediction to anomaly detection in cybersecurity. The scalability and efficiency of Transformers, particularly when trained on large datasets, have contributed to their widespread adoption, setting new benchmarks across multiple fields and paving the way for their use in OS fingerprinting and network security applications.

\paragraph{Tabular Transformers}
\label{subsubsubsec:tabular-transformers}

Tabular Transformers are an adaption of Transformers designed to process structured tabular data commonly found in CSV files, spreadsheets, and relational databases. By leveraging self-attention mechanisms, they effectively capture complex feature interactions, eliminating the need for extensive feature engineering. These models have demonstrated superior performance in classification, regression, and other predictive tasks across various domains \cite{gorishniy_revisiting_2023}.  

We selected this architecture because network traffic datasets in this field are typically structured as collections of network flows, inherently formatted as tabular data. Unlike traditional DL models such as CNNs and RNNs, which rely on spatial or sequential patterns, Tabular Transformers excel at modelling structured data with heterogeneous features. Their ability to learn intricate dependencies across multiple attributes makes them particularly well-suited for analysing network traffic.  

Furthermore, as will be exposed in Section~\ref{sec:related-work}, this approach has not yet been explored for OS fingerprinting, despite its potential to enhance classification accuracy by capturing nuanced relationships in network traffic data. This gap in prior research motivated our investigation into the applicability of Tabular Transformers to this task.

Several Transformer-based architectures have been proposed for tabular data modelling. For this study, we selected two representative models—TabTransformer (TabT) and FT-Transformer (FT-T)—due to their demonstrated effectiveness in handling categorical and numerical features, allowing us to assess their suitability for OS fingerprinting. In the context of network traffic analysis, categorical features represent discrete variables with distinct groups, such as protocol type or OS family, while numerical features are continuous values that quantify measurements, like packet size or TTL.

\begin{itemize}
    \item \textbf{TabTransformer (TabT)} \cite{huang_tabtransformer_2020} replaces traditional categorical embeddings with context-aware representations using Transformer layers. By applying multi-head self-attention to categorical features, it captures dependencies and interactions more effectively than standard embedding techniques, improving classification tasks and enhancing robustness to missing and noisy data.

    \item \textbf{FT-Transformer (FT-T)} \cite{gorishniy_revisiting_2023} generalizes the self-attention mechanism to both categorical and numerical features, treating them uniformly to better capture interactions across heterogeneous data. Unlike architectures that require extensive preprocessing or feature engineering, FT-Transformer learns feature dependencies directly from raw tabular inputs, making it well-suited for complex network traffic datasets with mixed feature types.
\end{itemize}

\begin{figure}[b!]
    \centering
    \includegraphics[width=0.74\textwidth]{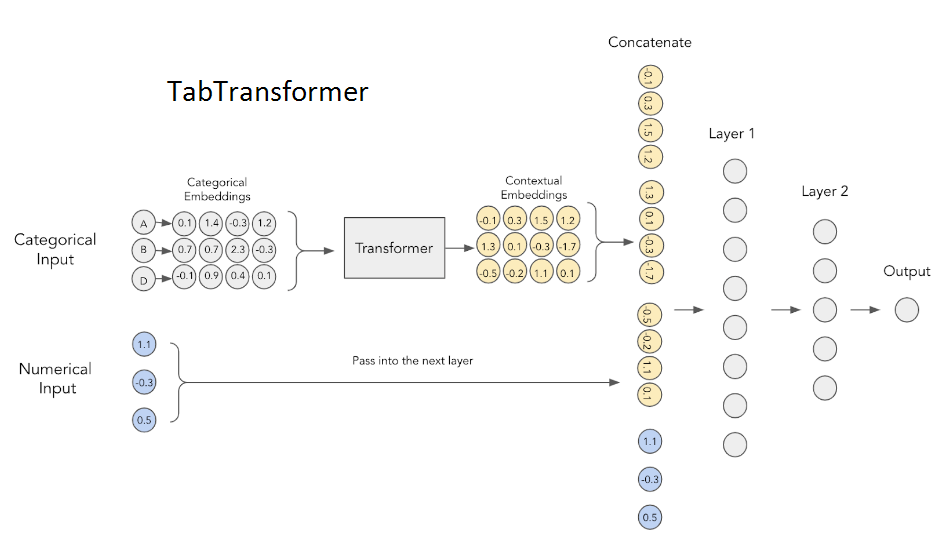}    
    \caption{Diagram of TabTransformer (Tab-T) architecture. Source: \cite{wang_lucidrainstab-transformer-pytorch_2024}}
    \label{fig:tabt}
\end{figure}

\begin{figure}[b!]
    \centering
    \includegraphics[width=0.74\textwidth]{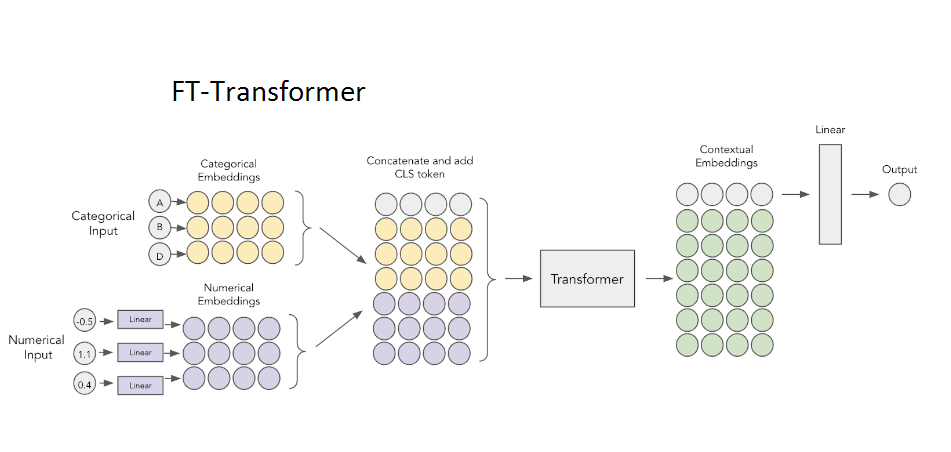}    
    \caption{Diagram of FT-Transformer (FT-T) architecture. Source: \cite{wang_lucidrainstab-transformer-pytorch_2024}}
    \label{fig:ftt}
\end{figure}

These models were selected to compare different feature integration strategies in OS fingerprinting. While TabTransformer processes only categorical variables through the Transformer, FT-Transformer applies self-attention to both categorical and numerical features. Figures \ref{fig:tabt}-\ref{fig:ftt} illustrate these differences, highlighting how each architecture structures and transforms input data for prediction tasks.

\clearpage

%% file: content/s03-related-work.tex
\section{Related Work}
\label{sec:related-work}

\subsection{OS Fingerprinting}
\label{subsec:soa-os-fingerprinting}

\subsubsection{Traditional OS Fingerprinting}
\label{subsubsec:traditional-os-fingerprinting}

OS fingerprinting initially emerged as an approach based solely on the analysis of TCP/IP header fields, such as Time To Live (TTL), the Don't Fragment (DF) flag, Type of Service (ToS) and TCP Window Size \cite{nmaporg_nmap_nodate, lastovicka_passive_2023}. In this context, Nmap was one of the first tools to be developed and remains one of the most widely used, as it employs an active scanning method that compels the target machine to respond, thereby facilitating the identification of its operating system.

Alternatively, passive fingerprinting techniques were subsequently proposed, relying on similar network traffic characteristics but without provoking a response from the target system. Early implementations of this approach, such as p0f and Siphon, emerged in 2000 \cite{zalewski_p0f_nodate, beddoe_siphon_nodate}. As previously discussed (Section~\ref{subsec:os-fingerprinting}), the fundamental difference between the two methods lies in the manner in which network information is collected, rather than in the analytical techniques applied to the traffic for inference.

Traditional OS fingerprinting methods based on TCP/IP headers continued to evolve. Tools like Ettercap and Satori extended earlier approaches \cite{ornaghi_alberto_ettercap_2001, e_kollmann_satori_2018}, while others like NetSleuth and PRADS saw limited longevity \cite{netgrab_netsleuth_2012, eb_fjellskal_prads_2009}. Research expanded into passive OS fingerprinting in large networks, examining features from network flows, as demonstrated by Vymlátil and Matoušek \cite{vymlatil_detection_2014, matousek_gromovs_2014}. These methods proved effective in dynamic environments such as wireless networks, and newer approaches—such as those by Al-Sherari and Osanaiye \cite{al-shehari_improving_2014, osanaiye_tcpip_2015}—combined traditional methods with ML to achieve better accuracy, particularly in unauthorized OS detection and cloud environments.

Modern approaches have shifted towards analysing application layer protocols, encrypted traffic, and specialised traffic types. Hypertext Transfer Protocol (HTTP) banners and User-Agent strings provide more precise OS identification \cite{shah_http_2003}, while encryption complicates traditional methods. Researchers like Muehlstein and Fan \cite{matousek_gromovs_2014, fan_identify_2019} improved accuracy by incorporating Transport Layer Security (TLS) handshake features, and others like Aksoy \cite{aksoy_operating_2016} explored various protocols using ML to optimise fingerprinting. In specific cases, such as smartphone OS identification and Industrial Control System (ICS) devices, timing analysis and advanced algorithms have been used \cite{gurary_operating_2016, shen_hybrid-augmented_2018}. ML has become essential in overcoming traditional limitations, focusing on processing large datasets and achieving higher accuracy in identifying OS in encrypted traffic \cite{beverly_robust_2004, shamsi_faulds_2021, lastovicka_cybersecurity_2018}. A detailed explanation of the works where ML is applied to the OS fingerprinting field is exposed in \Cref{sec:related-work}.

\subsubsection{Machine Learning-based OS Fingerprinting}
\label{subsubsec:ml-based-os-fingerprinting}

The field of OS fingerprinting has evolved significantly with the advent of ML techniques. In this context, several approaches have been proposed to enhance the accuracy and robustness of OS identification. For instance, Lastovicka's research \cite{lastovicka_cybersecurity_2018} explored various classical ML algorithms like Naive Bayes (NB), Decision Trees (DT), k-Nearest Neighbours (kNN), and Support Vector Machines (SVM) for passive OS fingerprinting. In a later study \cite{lastovicka_usingTLS_2020}, Lastovicka et al. expanded this work by employing TLS handshake features, which improved device identification even in encrypted network environments. Similarly, Fan et al. in \cite{fan_identify_2019} employed Gradient Boosting Decision Trees (GBDT) on features extracted from both TLS and TCP/IP headers, achieving high accuracy on a large dataset. This aligns with previous contributions made by our research team \cite{perez-jove_applying_2021, perez-jove_tool_2023}, where we applied a range of classical ML algorithms like NB, Multilayer Perceptron (MLP), DT, and Random Forest (RF) to the Nmap and p0f databases.

Other studies focused on minimalist data requirements and novel feature extraction. Millar et al. \cite{millar_operating_2020} used RF to classify OS types based on IP affiliation graphs, demonstrating resilience to encrypted traffic. Similarly, Barath et al. \cite{barath_use_2021} employed DT, Expectation-Maximization (EM), NB, and Artificial Neural Networks (ANNs) for passive monitoring, further showcasing the potential of various ML techniques for network data analysis. Shamsi et al. \cite{shamsi_faulds_2021} also used a non-parametric EM estimator to improve OS fingerprinting accuracy in noisy data for large-scale network environments and dealing with distortions.

Several works focused on specialised environments or different network protocols. Salah et al. \cite{salah_desktop_2022} focused on IPv6-based fingerprinting using kNN, DT, SVM, and Gaussian Naive Bayes (GNB), whereas Bub et al. \cite{bub_towards_2022} applied DT to identify aged Android devices in home networks. Hulák et al. \cite{hulak_evaluation_2023} compared the performance of DT, RF, and AdaBoost (AB) in passive OS fingerprinting, emphasising the importance of careful data preparation. In a related domain, Zhang et al. \cite{zhang_operating_2022} integrated Active Learning (AL) with SVM, RF, and NB to optimise OS identification, particularly in environments with dynamic network conditions.

The literature review efforts in this field are limited, with Lastovicka et al. \cite{lastovicka_passive_2023} providing the only comprehensive survey of passive OS fingerprinting methods, detailing the transition from traditional techniques to ML-based approaches. 

\subsubsection{Deep Learning-based OS Fingerprinting}
\label{subsubsec:dl-based-os-fingerprinting}

Even though ML techniques have been explored in OS fingerprinting, as previously outlined, there is little research on the application of DL models to this field. For the best of author's knowledge, only two works employed some DL algorithm to solve this network task. Li et al. \cite{li_passive_2023} proposed a combined sampling method paired with a Convolutional Neural Network (CNN) to improve identification accuracy for underrepresented OS types in imbalanced datasets. Hagos et al. \cite{hagos_machine-advanced_2020, hagos_TCPvariant_2021} introduced the TCP variant as a feature in ML models, exploring a mix of traditional ML algorithms, like SVM, RF, kNN, NB, with DL models like Long Short-Term Memory (LSTM). Finally, a preliminary version of this work was presented in \cite{perez_jove_towards_2024}, where the Transformer architecture was applied to the Nmap database.

\subsection{Transformers in Network Traffic Modelling}
\label{subsec:transformers-network-traffic}

Beyond its success in NLP, the Transformer architecture has proved versatile across domains such as computer vision (e.g., the Vision Transformer (ViT) \cite{han_survey_2023}) and network analysis. Although Transformers have not yet been applied directly to OS fingerprinting, they have been adapted for various networking tasks. For example, NetBERT outperforms BERT on network-specific tasks \cite{louis_netbert_2020}, while adaptations of BERT for DNS analysis and the use of Graph Neural Networks for packet sequences have also been explored \cite{le_norbert_2022}. Direct training on network traffic has yielded promising results too, with the Residual 1-D Image Transformer excelling in malware and DDoS detection \cite{barut_r1dit_2023} and the Flow Transformer enhancing anonymity network classification by capturing temporal–spatial dependencies \cite{zhao_flow_2021}. Furthermore, De la Torre Vico et al. \cite{de_la_torre_vico_exploring_2024} have demonstrated the potential of LLMs in analysing network traces for cybersecurity.

Concurrently, interest in Network Traffic Foundational Models (NT-FMs) inspired by large language models is growing. Early work includes ET-BERT, which leverages contextualised datagram representations for encrypted traffic classification \cite{lin_et-bert_2022}, and Ray’s packet-level traffic prediction model \cite{ray_advancing_2022}. Subsequent studies have investigated model generalisation \cite{dietmuller_new_2022} and foundational applications in networking \cite{le_rethinking_2022}. More recent advances include Zhao et al.’s Yet Another Traffic Classifier (YATC) using a masked autoencoder with multi-level flow representations \cite{zhao_yet_2023}, Guthula et al.’s netFound utilising unlabelled traffic for pre-training \cite{guthula_netfound_2023}, and Wang et al.’s Lens capturing temporal–spatial correlations for anomaly detection \cite{wang_lens_2024}. Additional contributions include TrafficGPT for long traffic sequence modelling \cite{qu_trafficgpt_2024}, a graph-based NT-FM by Langendonck et al. for improved scalability \cite{van_langendonck_towards_2024}, the generative pretrained model NetGPT for traffic understanding and generation \cite{meng_netgpt_2023}, and the comprehensive NetBench dataset for evaluating foundational models on traffic tasks \cite{qian_netbench_2024}.

%% file: content/s04-materials-methods.tex
\section{Materials \& Methods}
\label{sec:materials-methods}


\subsection{Datasets}
\label{subsec:datasets}

For evaluating OS fingerprinting methods, we selected publicly available benchmark datasets that meet essential criteria for evaluating OS fingerprinting methods. Specifically, datasets must have sufficient size, diversity, and accurate OS labelling, and given the rapid evolution of OSs, they need to be up-to-date. Our selection process was informed by their adoption in recent studies and an analysis of prior works (Section~\ref{sec:related-work}), ensuring both relevance and comparability. Moreover, these datasets capture network traffic at various abstraction levels—from high-level flow summaries to detailed packet-level data—and provide OS labels at multiple granular levels (family, major, and minor versions).  

We selected three representative datasets—\texttt{DAT1}, \texttt{DAT2}, and \texttt{DAT3}—that capture diverse network features essential for robust OS fingerprinting. In our review of prior works, we analysed available features like packets, telemetry, and logs to identify datasets that are both recent and reflective of realistic network environments with current OS versions and modern traffic dynamics. These datasets encompass diverse data types (flow-level records, packet captures, and active OS fingerprinting signatures), offer multiple levels of OS granularity (family, major, and minor versions), vary in size, and originate from different network settings. This diversity underpins a robust evaluation of AI-based OS fingerprinting models by mitigating dataset bias and enhancing generalisability.

Three complementary data types—IPFIX, PCAP, and OS signature databases— were employed to comprehensively analyse network traffic for OS fingerprinting. Specifically, IPFIX (Internet Protocol Flow Information Export) provides flow-level metadata summarizing network activity; PCAP (Packet CAPture) retains raw network packets to capture detailed protocol behavior; and OS signature databases, such as those used by Nmap, offer predefined OS fingerprints as classification references. Together, these sources enable a multifaceted analysis that integrates both high-level traffic patterns and in-depth protocol details.

\clearpage

We defined OS classification tasks at three granularity levels—\texttt{family}, \texttt{major}, and \texttt{minor}—to capture varying levels of detail. Specifically, the \texttt{family} level includes broad categories (e.g., Windows, Linux, Android), the \texttt{major} level distinguishes versions (e.g., Windows 10, Android 9, iOS 13), and the \texttt{minor} level provides finer distinctions (e.g., Windows 8.1, iOS 13.5, macOS 10.15). For instance, a detailed classification might separate \texttt{Ubuntu} into \texttt{22.04 (Jammy Jellyfish)} and further into \texttt{22.04.3 LTS}, whereas a less detailed approach would label it simply as \texttt{Ubuntu}. Higher granularity, particularly at the \texttt{minor} level, increases classification complexity due to a larger number of classes and fewer training examples per class. The characteristics of the selected datasets are further detailed in the following points and summarized in \Cref{tab:datasets}.

\begin{table}[t!]
\centering
\caption{Overview of the employed datasets}
\label{tab:datasets}
\begin{tblr}{
  width = \linewidth,
  colspec = {Q[50]Q[37]Q[40]Q[54]Q[38]Q[50]Q[35]Q[35]Q[35]Q[35]Q[65]Q[75]Q[50]},
  cells = {c},
  row{4} = {my_grey}, 
  row{5} = {my_grey}, 
  row{6} = {my_grey}, 
  cell{1}{1} = {r=2}{},
  cell{1}{2} = {r=2}{},
  cell{1}{3} = {r=2}{},
  cell{1}{4} = {r=2}{},
  cell{1}{5} = {c=6}{0.247\linewidth},
  cell{1}{11} = {r=2}{},
  cell{1}{12} = {r=2}{},
  cell{1}{13} = {r=2}{},
  cell{4}{1} = {r=3}{},
  cell{4}{2} = {r=3}{},
  cell{4}{3} = {r=3}{},
  cell{4}{4} = {r=3}{},
  cell{4}{5} = {r=3}{},
  cell{4}{6} = {r=3}{},
  cell{4}{7} = {r=3}{},
  cell{4}{8} = {r=3}{},
  cell{4}{9} = {r=3}{},
  cell{4}{10} = {r=3}{},
  cell{4}{11} = {r=3}{},
  cell{7}{1} = {r=2}{},
  cell{7}{2} = {r=2}{},
  cell{7}{3} = {r=2}{},
  cell{7}{4} = {r=2}{},
  cell{7}{5} = {r=2}{},
  cell{7}{6} = {r=2}{},
  cell{7}{7} = {r=2}{},
  cell{7}{8} = {r=2}{},
  cell{7}{9} = {r=2}{},
  cell{7}{10} = {r=2}{},
  cell{7}{11} = {r=2}{},
  hline{1,3,9} = {-}{},
}
\textbf{Dataset} & \textbf{Year} & \textbf{Works} & \textbf{Data Type} & \textbf{Feature Count} &  &  &  &  &  & \textbf{Row Count} & \textbf{Granularity} & \textbf{Classes Count}\\
 &  &  &  & \textbf{Total} & \textbf{TCP/IP} & \textbf{DNS} & \textbf{HTTP} & \textbf{TLS} & \textbf{Other} &  &  & \\
\texttt{DAT1} \cite{martin_dataset_2019} & 2019 & \cite{lastovicka_usingTLS_2020} & IPFIX & 29 & 7 & - & 5 & 8 & 9 & 18,708,983 & \texttt{family} & 5\\
\texttt{DAT2} \cite{lastovicka_dataset_2023} & 2023 & \cite{lastovicka_passive_2023, hulak_evaluation_2023} & IPFIX & 112 & 35 & - & 7 & 28 & 46 & 109,663 & \texttt{family} & 12\\
 &   &   &   &   &   &   &   &   &   &   & \texttt{major} & 50\\
  &   &   &   &   &   &   &   &   &   &   & \texttt{minor} & 88\\
\texttt{DAT3} \cite{noauthor_nmap794-osdb} & 2023 & \cite{perez-jove_applying_2021, perez_jove_towards_2024} & DB & 263 & 263 & - & - & - & - & 38,817 & \texttt{family} & 7\\
 &  &  &  &  &  &  &  &  &  &  & \texttt{minor} & 91
\end{tblr}
\end{table}

\begin{itemize}

\item \texttt{\textbf{DAT1}}: \texttt{\textbf{lastovicka\_2019\_UsingTLS}} \cite{martin_dataset_2019}  

This dataset comprises flow records from the Czech Republic Masaryk university's backbone network, enriched with log entries from Dynamic Host Configuration Protocol (DHCP) servers and a Remote Authentication Dial-In User Service (RADIUS) accounting server. Collected between July 12 and 16, 2019, it focuses on flows originating from the university's Eduroam wireless networks. OS labels are derived from DHCP logs and RADIUS session IDs.  

Useful features for OS fingerprinting include basic flow attributes, extended TCP/IP parameters, HTTP user-agent strings, and TLS client details. The dataset, anonymized using the Crypto-PAn algorithm, spans 18.7 million rows with 29 features, primarily supporting OS family-level classification with five classes.

\item \texttt{\textbf{DAT2}}: \texttt{\textbf{lastovicka\_2023\_PassiveOSRevisited}} \cite{lastovicka_dataset_2023}  

This dataset captures web traffic from five Masaryk university servers hosting 475 domains over eight hours. OS labels are derived from HTTP User-Agent strings in web server logs, cross-referenced with network flow data. Collected connections include devices such as user computers, mobile phones, and web crawlers.  

OS fingerprinting-related features include IP and TCP parameters, HTTP and TLS details, among others, amounting to 112 features in total. The dataset includes 109,663 rows, enabling OS classification at family (12 classes), major (50 classes), and minor (88 classes) levels.

\item \texttt{\textbf{DAT3}}: \texttt{\textbf{nmap-7.94\_2023\_OSdb}} \cite{noauthor_nmap794-osdb}  

This dataset consists of OS signatures actively collected using the Nmap tool (version 7.94). Nmap identifies OSs by sending 16 TCP, User Datagram Protocol (UDP), and Internet Control Message Protocol (ICMP) probes and analysing responses. Features include TCP window sizes, sequence generation, and TCP options, providing detailed fingerprinting data.  

The dataset contains 38,817 rows with 263 features and supports fine-grained OS classification at family (7 classes) and minor (91 classes) levels. Its active collection process complements the passive data in \texttt{DAT1} and \texttt{DAT2} by capturing precise protocol behaviours.

\end{itemize}

\begin{figure}[t!]
    \centering
    \begin{subfigure}{0.49\textwidth}
        \centering
        \includegraphics[width=\linewidth]{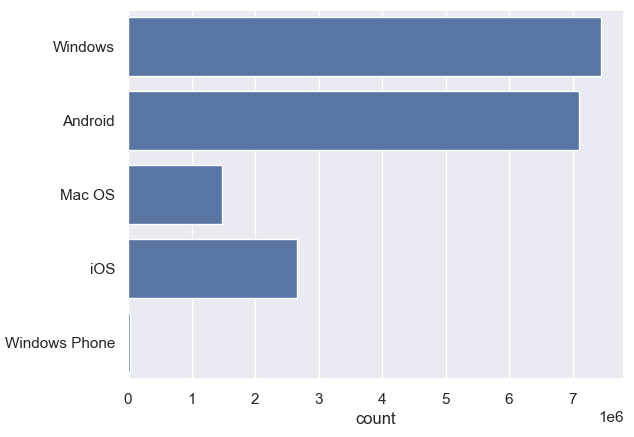}
        \caption{\texttt{DAT1}}
        \label{fig:class-distribution-1a}
    \end{subfigure}
    \hfill
    \begin{subfigure}{0.49\textwidth}
        \centering
        \includegraphics[width=\linewidth]{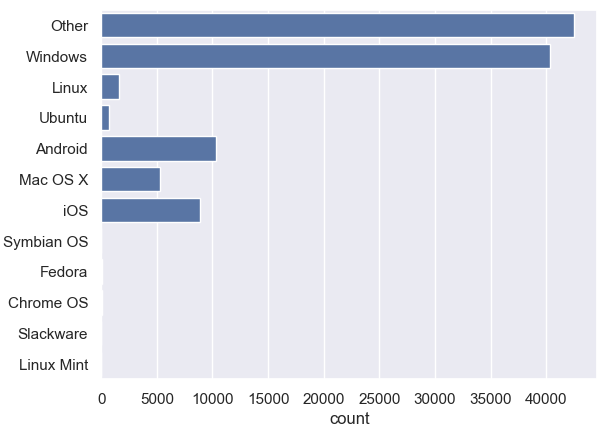}
        \caption{\texttt{DAT2}}
        \label{fig:class-distribution-1b}
    \end{subfigure}

    \begin{subfigure}{0.5\textwidth}
        \centering
        \includegraphics[width=\linewidth]{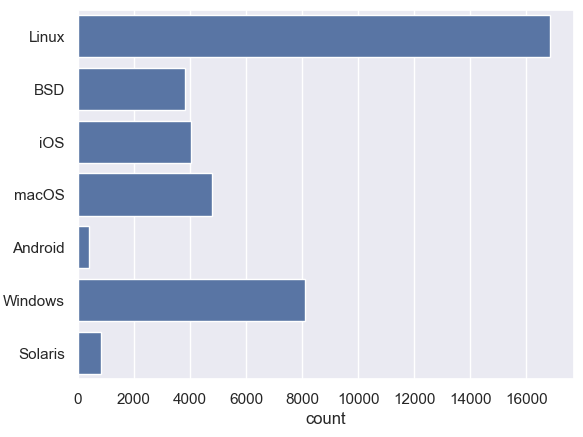}
        \caption{\texttt{DAT3}}
        \label{fig:class-distribution-1c}
    \end{subfigure}

    \caption{Classes distribution by OS \texttt{family} version in \texttt{DAT1} (\textit{top-left}), \texttt{DAT2} (\textit{top-right}), and \texttt{DAT3} (\textit{bottom-centre}).}
    \label{fig:class-distribution}
\end{figure}

Class distribution analysis revealed significant imbalances across all datasets, meaning that some classes contain far more examples than others. This imbalance, illustrated in \Cref{fig:class-distribution}, introduces challenges for training, as models may become biased toward majority classes. Addressing this imbalance is critical to achieving robust and fair performance across all tasks, and will be discussed in Section~\ref{subsec:data-preparation}.

\clearpage

\subsection{Data Preparation}
\label{subsec:data-preparation}

Data preparation is the process of cleaning, transforming, and balancing datasets to ensure robust model performance and reproducibility. In our study, this involves handling missing or invalid data, removing irrelevant features, and addressing class imbalances—each with fixed parameters and random seeds to guarantee full reproducibility.

\paragraph{Handling Missing Data and Redundancies.} We improve data quality by systematically addressing missing values and redundant entries. In the datasets, Not a Number (NaN) or Null values—indicating undefined or unrepresentable quantities (e.g., missing TLS features in unencrypted flows)—are encoded categorically when applicable; rows with missing values in critical columns are removed, and numerical features with zero variance (i.e., variance $\leq 0$) are dropped. Duplicate entries are also removed to avoid redundancy, and no other errors or infinite values were detected.

\paragraph{Removing Irrelevant Features.} Irrelevant columns are removed to focus on OS fingerprinting. For example, columns related to timestamps or non-OS network information—such as \texttt{Date flow start} and \texttt{Session ID} in \texttt{DAT1}—are excluded, with similar removals performed in \texttt{DAT2} and \texttt{DAT3}. The precise selection of retained features is provided in Table~\ref{tab:selected-features}, along with explicit lists of both categorical and numerical features to standardize the preprocessing pipeline.

\paragraph{Addressing Class Imbalances.} We mitigate class imbalances by standardizing target classes and applying resampling techniques. Target classes are first standardized via regular expression matching (e.g., mapping entries containing 'iOS', 'Android', 'Mac OS X', and 'Windows' to their respective labels). Then, random undersampling is applied to majority classes using predetermined removal percentages (with fixed random seeds), while the Synthetic Minority Over-sampling Technique (SMOTE) is employed with a sampling strategy of \texttt{'auto'} to generate synthetic samples for minority classes:
\begin{enumerate}
    \item \textbf{Random Undersampling:} Applied to majority classes when sufficient data is available.
    \item \textbf{SMOTE:} Employed to generate synthetic samples for minority classes.
\end{enumerate}

\paragraph{Feature Processing and Encoding.} Feature processing further refines both categorical and numerical data for improved interpretability and balanced training. Categorical columns containing hexadecimal strings are split into individual bytes, and One-Hot Encoding is applied to categorical target variables—converting them into a numerical format by creating binary columns for each class, after which the original target column is removed. Additionally, class weights are computed as the inverse of normalized class frequencies, and data is split into training and test sets using stratified sampling (with 20\% reserved for testing).

\paragraph{Reproducibility and Software Environment.} Reproducibility is ensured by using fixed versions of key libraries and by sharing the complete code. Our experiments rely on \texttt{numpy==1.23.0}, \texttt{pandas==2.2.2}, \texttt{scikit-learn==1.5.0}, \texttt{torch==2.3.1}, and \texttt{optuna==3.6.1} for data manipulation, model training, and hyperparameter optimization. The full code implementing these steps is available at \url{https://github.com/rubenpjove/tabularT-OS-fingerprinting}.

\begin{table}[t!]
  \centering
  \caption{Selected Features for Each Dataset}
  \vspace{5pt}
  \begin{tabularx}{\linewidth}{>{\centering\arraybackslash}m{1cm} >{\centering\arraybackslash}X >{\centering\arraybackslash}X}
    \hline
    \textbf{Dataset} & \textbf{Numerical Features} & \textbf{Categorical Features} \\ \hline
    \texttt{DAT1} & {\small \texttt{TLS Client Version}, \texttt{Client Cipher Suites}, \texttt{TLS Extension Types}, \texttt{TLS Extension Lengths}, \texttt{TLS Elliptic Curves}, \texttt{TLS EC Point Formats}} & {\small \texttt{SYN size}, \texttt{TCP win}, \texttt{TCP SYN TTL}} \\ 
    \rowcolor{my_grey} \texttt{DAT2} & {\small \texttt{TCP flags A}, \texttt{TLS\_CONTENT\_TYPE}, \texttt{TLS\_HANDSHAKE\_TYPE}, \texttt{TLS\_CIPHER\_SUITE}, \texttt{TLS\_CLIENT\_VERSION}, \texttt{TLS\_CIPHER\_SUITES}, \texttt{TLS\_CLIENT\_SESSION\_ID}, \texttt{TLS\_EXTENSION\_TYPES}, \texttt{TLS\_CLIENT\_KEY\_LENGTH}, \texttt{TLS\_EXTENSION\_LENGTHS}, \texttt{TLS\_ELLIPTIC\_CURVES}, \texttt{TLS\_EC\_POINT\_FORMATS}, \texttt{IPv4DontFragmentforward}, \texttt{tcpOptionWindowScaleforward}, \texttt{tcpOptionSelectiveAckPermittedforward}, \texttt{tcpOptionNoOperationforward}, \texttt{flowEndReason}, \texttt{TLS\_JA3\_FINGERPRINT}, \texttt{IP ToS}} & {\small \texttt{SRC port}, \texttt{TCP SYN Size}, \texttt{TCP Win Size}, \texttt{TCP SYN TTL}, \texttt{NPM\_CLIENT\_NETWORK\_TIME}, \texttt{NPM\_ROUND\_TRIP\_TIME}, \texttt{NPM\_RESPONSE\_TIMEOUTS\_A}, \texttt{NPM\_TCP\_RETRANSMISSION\_A}, \texttt{NPM\_TCP\_OUT\_OF\_ORDER\_A}, \texttt{NPM\_JITTER\_DEV\_A}, \texttt{NPM\_JITTER\_AVG\_A}, \texttt{NPM\_JITTER\_MIN\_A}, \texttt{NPM\_JITTER\_MAX\_A}, \texttt{NPM\_DELAY\_DEV\_A}, \texttt{NPM\_DELAY\_AVG\_A}, \texttt{NPM\_DELAY\_MIN\_A}, \texttt{NPM\_DELAY\_MAX\_A}, \texttt{NPM\_DELAY\_HISTOGRAM\_1\_A}, \texttt{TLS\_SETUP\_TIME}, \texttt{tcpOptionMaximumSegmentSizeforward}} \\ 
    \texttt{DAT3} & {\small \texttt{SEQ.SP}, \texttt{SEQ.GCD}, \texttt{SEQ.ISR}, \texttt{SEQ.TI}, \texttt{SEQ.CI}, \texttt{SEQ.II}, \texttt{SEQ.TS}, \texttt{WIN.W*}, \texttt{ECN.T}, \texttt{ECN.TG}, \texttt{ECN.W}, \texttt{T*.T}, \texttt{T*.TG}, \texttt{T*.RD}, \texttt{T*.W}, \texttt{U1.T}, \texttt{U1.TG}, \texttt{U1.IPL}, \texttt{U1.UN}, \texttt{U1.RIPL}, \texttt{U1.RID}, \texttt{U1.RUCK}, \texttt{IE.T}, \texttt{IE.TG}, \texttt{IE.CD}} & {\small \texttt{SEQ.TI}, \texttt{SEQ.CI}, \texttt{SEQ.II}, \texttt{SEQ.SS}, \texttt{SEQ.TS}, \texttt{OPS.O1}, \texttt{OPS.O2}, \texttt{OPS.O3}, \texttt{OPS.O4}, \texttt{OPS.O5}, \texttt{OPS.O6}, \texttt{ECN.R}, \texttt{ECN.DF}, \texttt{ECN.O}, \texttt{ECN.CC}, \texttt{ECN.Q}, \texttt{T*.R}, \texttt{T*.DF}, \texttt{T*.S}, \texttt{T*.A}, \texttt{T*.F}, \texttt{U1.R}, \texttt{U1.DF}, \texttt{U1.RIPL}, \texttt{U1.RID}, \texttt{U1.RIPCK}, \texttt{U1.RUCK}, \texttt{U1.RUD}, \texttt{IE.R}, \texttt{IE.DFI}, \texttt{IE.CD}} \\ \hline
  \end{tabularx}
  \label{tab:selected-features}
\end{table}


\subsection{Modelling}
\label{subsec:modelling}

We present a comprehensive, reproducible, and efficient approach to model selection, training, and validation. After preparing each dataset, we first split the data into training and testing sets using stratified sampling to maintain class distribution—this same strategy is applied within cross-validation splits. The training set is used exclusively for model development, while the test set is reserved for final performance evaluation. To obtain optimal hyperparameters for the TabTransformer (TabT) \cite{huang_tabtransformer_2020} and FT-Transformer (FT-T) \cite{gorishniy_revisiting_2023}, we conducted a hyperparameter search using Optuna's NSGA-II sampler (Table~\ref{tab:hyperparameters}). In each trial, key metrics such as training time, inference time, model parameter count, and memory usage were recorded as user attributes to enhance reproducibility and analysis. Notably, all random seeds are fixed to guarantee reproducibility, and our approach is computationally more efficient than traditional methods like Grid or Random Search.

\begin{table}[t!]
\centering
\caption{Hyperparameters values}
\vspace{5pt}
\begin{tabular}{ccp{6cm}}
\hline
\textbf{Hyperparameter}       & \textbf{Value} & \textbf{Description} \\ \hline
\texttt{learning\_rate}                 & [0.0001 - 0.1]                                      & The range of learning rates used during training. \\ 
{\cellcolor{my_grey}}\texttt{embedding\_dim}                 & {\cellcolor{my_grey}}[16, 32]                                            & {\cellcolor{my_grey}}The dimensionality of the embeddings for categorical features. \\ 
\texttt{depth}                         & [2 - 6]                                             & The range for the number of Transformer layers. \\ 
{\cellcolor{my_grey}}\texttt{heads}                         & {\cellcolor{my_grey}}[2 - 8]                                             & {\cellcolor{my_grey}}The range for the number of attention heads in each transformer layer. \\ 
\texttt{attn\_dropout}                  & [0.05 - 0.5]                                        & The range for the dropout rate for the attention mechanism. \\ 
{\cellcolor{my_grey}}\texttt{ff\_dropout}                    & {\cellcolor{my_grey}}[0.05 - 0.5]                                        & {\cellcolor{my_grey}}The range for the dropout rate for the feedforward network within each Transformer layer. \\
\texttt{use\_shared\_categ\_embed}       & [True, False]                 & Determines whether to use shared embeddings for categorical features. Applicable to TabTransformer only.\\ \hline
\end{tabular}%
\label{tab:hyperparameters}
\end{table}

\paragraph{Hyperparameter Search Trials.} In each trial, we employ Stratified 10-Fold Cross-Validation with resampling techniques to handle class imbalance and optimize model performance. Specifically, data is split into training and validation sets with balanced class distributions across all folds. For class imbalance, random undersampling (with explicit removal percentages for majority classes) and SMOTE (with a fixed random state) are applied. Each fold undergoes 200 training epochs with batch sizes of 128, 256, or 512, and early stopping is triggered after 15 epochs without improvement. Model parameters are updated using the AdamW optimizer and training is guided by Cross-Entropy Loss. This entire process is executed on a Compute Unified Device Architecture (CUDA)-enabled device, utilizing data parallelization and custom Graphics Processing Unit (GPU) memory management (in conjunction with garbage collection) to ensure efficient computations and prevent memory leaks.

\paragraph{Baseline Models.} We also trained several baseline models—k-Nearest Neighbors (kNN), Random Forest (RF), and Multi-layer Perceptron (MLP)—to provide a comparative performance analysis. For these models, the same preprocessing pipeline was applied, except that categorical features were encoded using One-Hot Encoding. Each baseline model was trained with its default hyperparameters (see \autoref{tab:baselines_hyp}) and evaluated using Stratified 10-Fold Cross-Validation, with performance metrics averaged across folds. Detailed descriptions of these baseline models are provided in Section~\ref{subsubsec:ML-baselines}.

\begin{table}[h!]
    \centering
    \caption{Default hyperparameters for baseline ML models}
    \vspace{5pt}
    \begin{tabular}{cc}
    \hline
    \textbf{Model}       & \textbf{Hyperparameters}     \\ \hline
    \multirow{5}{*}{kNN} & \texttt{algorithm}: auto              \\
                         & {\cellcolor{my_grey}}\texttt{leaf\_size}: 30               \\
                         & \texttt{metric}: minkowski            \\
                         & {\cellcolor{my_grey}}\texttt{n\_neighbors}: 5              \\
                         & \texttt{weights}: uniform             \\ \hline
    \multirow{6}{*}{RF}  & \texttt{n\_estimators}: 100           \\
                         & {\cellcolor{my_grey}}\texttt{criterion}: gini              \\
                         & \texttt{max\_depth}: None             \\
                         & {\cellcolor{my_grey}}\texttt{min\_samples\_split}: 2       \\
                         & \texttt{min\_samples\_leaf}: 1        \\
                         & {\cellcolor{my_grey}}\texttt{random\_state}: 42            \\ \hline
    \multirow{6}{*}{MLP} & \texttt{hidden\_layer\_sizes}: (100,) \\
                         & {\cellcolor{my_grey}}\texttt{activation}: relu             \\
                         & \texttt{solver}: adam                 \\
                         & {\cellcolor{my_grey}}\texttt{alpha}: 0.0001                \\
                         & \texttt{learning\_rate}: constant     \\
                         & {\cellcolor{my_grey}}\texttt{random\_state}: 42            \\ \hline
    \end{tabular}
    \label{tab:baselines_hyp}
\end{table}

\clearpage

\subsubsection{Computational Resources}
\label{subsubsec:hardware}

This research leveraged the FinisTerrae III supercomputer at CESGA \cite{ft3} for model training. This systems is a Bull ATOS bullx configured across 13 racks, which includes 714 Intel Xeon Ice Lake 8352Y processors and 157 GPUs (141 Nvidia A100 and 16 Nvidia T4 units). It has 126 TB of memory, 359 TB of SSD NVMe storage, and Infiniband HDR 100 for networking, achieving a peak performance of 4.36 PetaFLOPS. Different hardware configurations were used for the experiments, based on the computational requirements of each task and the availability within the system's job scheduling system.


\subsection{Evaluation}
\label{subsec:evaluation}

This section details our comprehensive evaluation framework designed to rigorously assess the performance and efficiency of the proposed Tabular Transformer models for OS fingerprinting.

After hyperparameter optimization, the best-performing Tabular Transformer model was retrained on the full training dataset and evaluated on a hold-out test set using the same preprocessing pipeline. Early stopping was employed during training to prevent overfitting, and all splits were generated via stratified sampling to preserve the class distribution inherent to the OS fingerprinting task.

The evaluation metrics included are accuracy, precision, recall, and F1-score. For the latest three, as we are evaluating a multiclass classification problem, we employed the weighted average technique. These metrics, defined in the following Subsection~\ref{subsubsec:metric-definitions}, provide a robust assessment of the methods' performance. The accuracy metric itself can be very misleading on imbalanced datasets where one target class dominates the dataset. Furthermore, as when it comes to evaluate OS fingerprinting, we want to have a good balance between precision (how accurate the model is in its positive predictions) and recall (how complete the model's positive predictions are). Therefore, the metric we want to focus when comparing different results is F1-score, which is an harmonic mean of both.

In addition to accuracy metrics, we recorded the total training time and measured inference time, while also computing key model characteristics such as the number of trainable parameters and the overall memory footprint.

Furthermore, confusion matrices were generated to analyse misclassification at the class level. Final predictions along with their corresponding ground truth labels were saved to a CSV file, and the list of class labels was written to a separate text file. These steps ensure that the evaluation results are fully reproducible and facilitate subsequent analyses of the model's performance on an imbalanced, multiclass OS fingerprinting problem.

\subsubsection{Evaluation Metrics Definitions}
\label{subsubsec:metric-definitions}

Our evaluation relies on the following metrics:

\paragraph{Accuracy} measures the overall correctness of predictions:
\[
\text{Accuracy} = \frac{TP + TN}{TP + TN + FP + FN}
\]

\paragraph{Balanced Accuracy} computes the average recall per class, ensuring equal contribution from all classes:
\[
\text{Balanced Accuracy} = \frac{1}{C} \sum_{i=1}^{C} \frac{TP_i}{TP_i + FN_i}
\]
where \(C\) is the number of classes.

\paragraph{Precision} quantifies the proportion of true positives among all positive predictions:
\[
\text{Precision} = \frac{TP}{TP + FP}
\]

\paragraph{Recall} (Sensitivity) measures the proportion of true positives identified among all actual positives:
\[
\text{Recall} = \frac{TP}{TP + FN}
\]

\paragraph{F1-Score} is the harmonic mean of Precision and Recall:
\[
\text{F1-Score} = 2 \cdot \frac{\text{Precision} \cdot \text{Recall}}{\text{Precision} + \text{Recall}}
\]

\paragraph{Weighted Averages} are computed to account for class imbalance by weighting each class metric by its support:
\[
\text{Weighted Metric} = \frac{\sum_{i=1}^{C} (\text{Support}_i \times \text{Metric}_i)}{\sum_{i=1}^{C} \text{Support}_i}
\]

%% file: content/s05-results-discussion.tex
\section{Results \& Discussion}
\label{sec:results-discussion}

Our experiments demonstrate that Transformer-based models, particularly FT-Transformer, generally outperform classical ML methods in OS fingerprinting tasks across various datasets and classification granularities. In this work, we applied two Transformer architectures—TabTransformer (TabT) and FT-Transformer (FT-T)—to three distinct datasets featuring OS information at different levels (from broad family categories to detailed major and minor versions) and compared their performance with traditional models (kNN, RF, MLP).

\paragraph{Hyperparameter Optimization.} Optimal hyperparameters for each Tabular Transformer architecture were identified (see \autoref{tab:best-hyps}). Using these parameters, we trained the models and evaluated them on the different datasets and classification tasks.

\begin{itemize}
    \item \textbf{\texttt{DAT1}:} Only OS \texttt{family} classification was feasible. FT-Transformer significantly outperformed TabTransformer and traditional models (kNN, RF, MLP) in all computed metrics, as shown in \autoref{tab:metrics-DAT1}.
    
    \item \textbf{\texttt{DAT2}:} This dataset supported \texttt{family}, \texttt{major}, and \texttt{minor} classifications. For \texttt{family} classification, the Random Forest (RF) model marginally outperformed the Transformer-based models; however, for both \texttt{major} and \texttt{minor} levels, FT-Transformer achieved the highest performance (see \autoref{tab:metrics-DAT2}).
    
    \item \textbf{\texttt{DAT3}:} With \texttt{family} and \texttt{minor} classifications available, FT-Transformer led in the \texttt{family} category, while TabTransformer showed superior performance for the \texttt{minor} level (refer to \autoref{tab:metrics-DAT3}).
\end{itemize}

\begin{figure}[]
    \centering
    \includegraphics[width=0.9\textwidth]{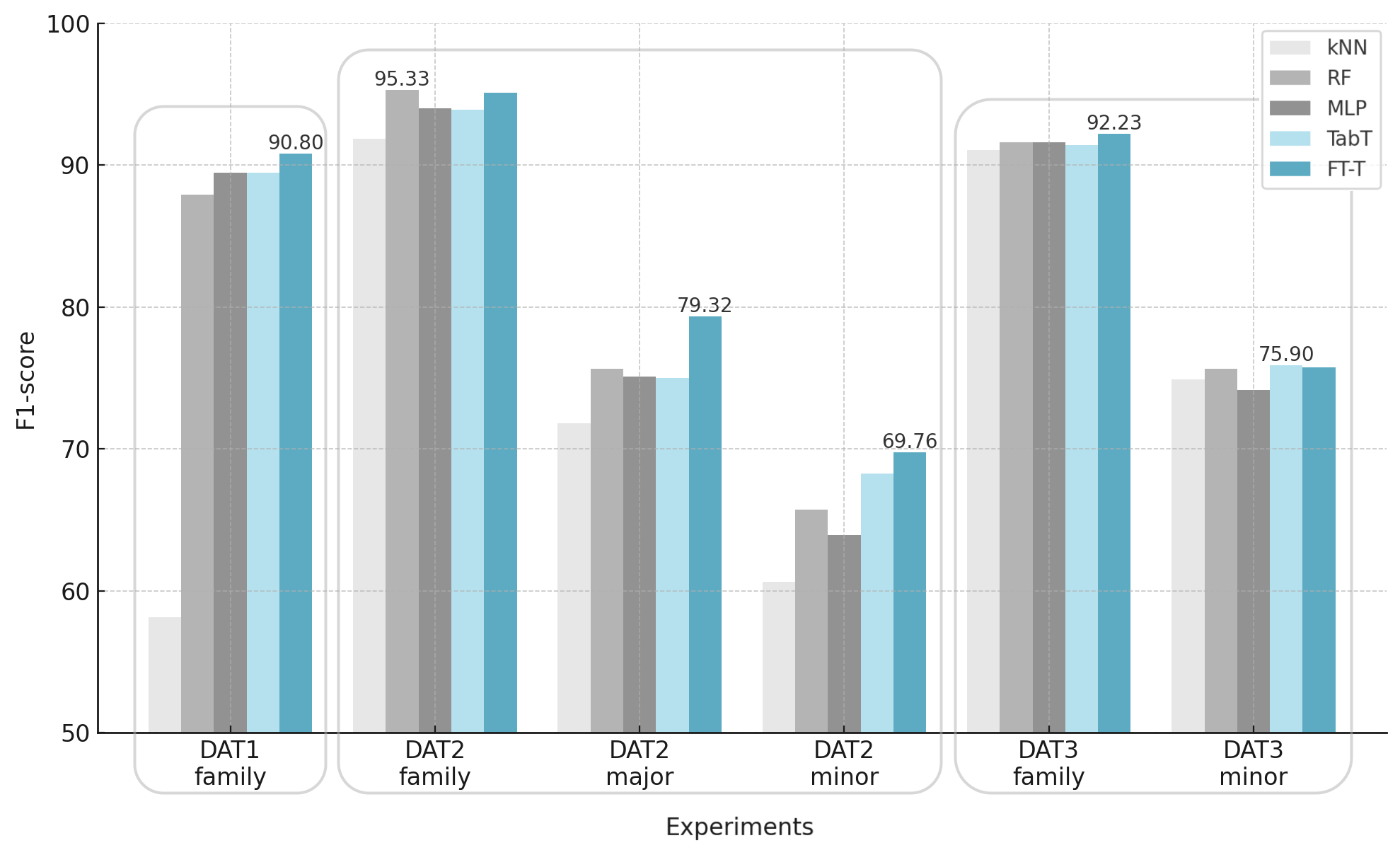}
    \caption{Comparison of F1-scores (weighted) across tested models for various datasets and classification tasks. Each experiment represents a dataset and granularity of the classification, with the highest performance for each experiment annotated at the top of the corresponding bar}
    \label{fig:results}
\end{figure}

\begin{figure}[]
    \centering
    \begin{subfigure}{0.42\textwidth}
        \centering
        \includegraphics[width=\linewidth]{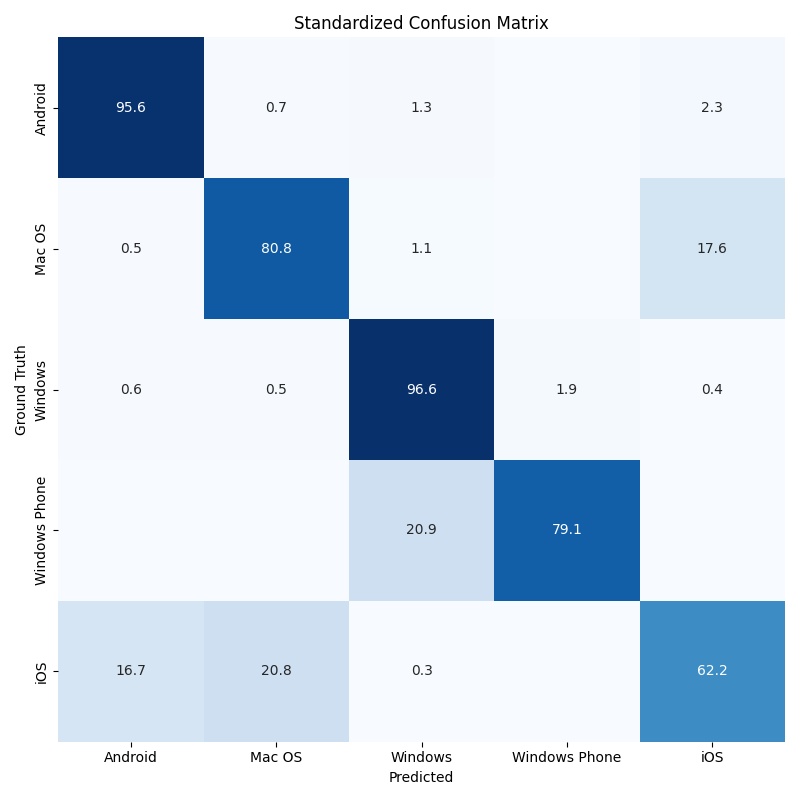}
        \label{fig:conf-matrix-1a}
    \end{subfigure}
    \begin{subfigure}{0.42\textwidth}
        \centering
        \includegraphics[width=\linewidth]{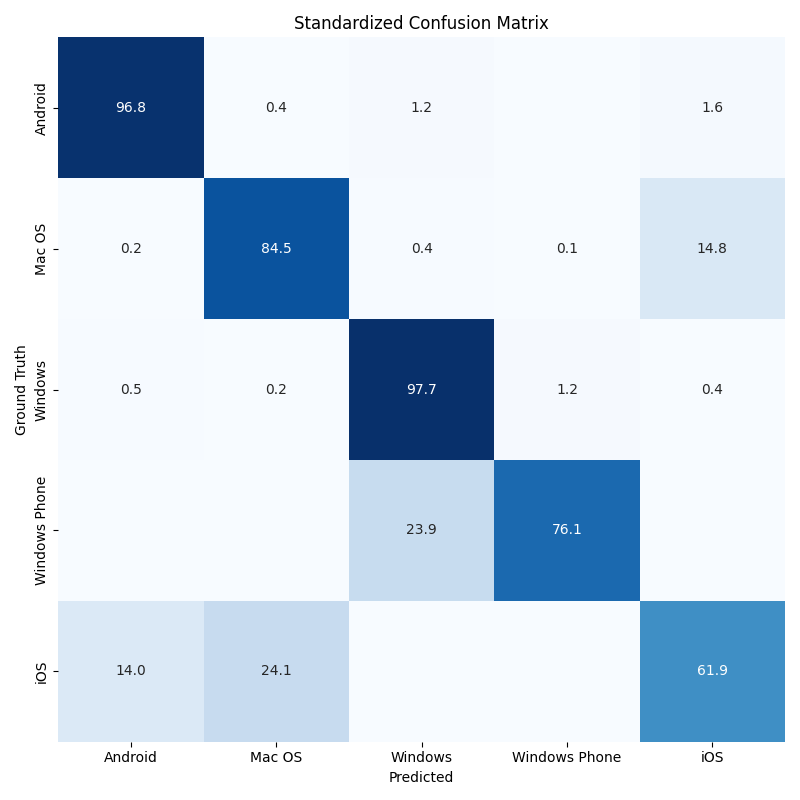}
        \label{fig:conf-matrix-1b}
    \end{subfigure}
    \vspace{-15pt}
    \caption{Standardised confusion matrices for the \texttt{family} classification in \texttt{DAT1} (TabT: \textit{left}, FT-T: \textit{right})}
    \label{fig:conf-matrix-dat1}
\end{figure}

\begin{figure}[]
    \centering
    \begin{subfigure}{0.42\textwidth}
        \centering
        \includegraphics[width=\linewidth]{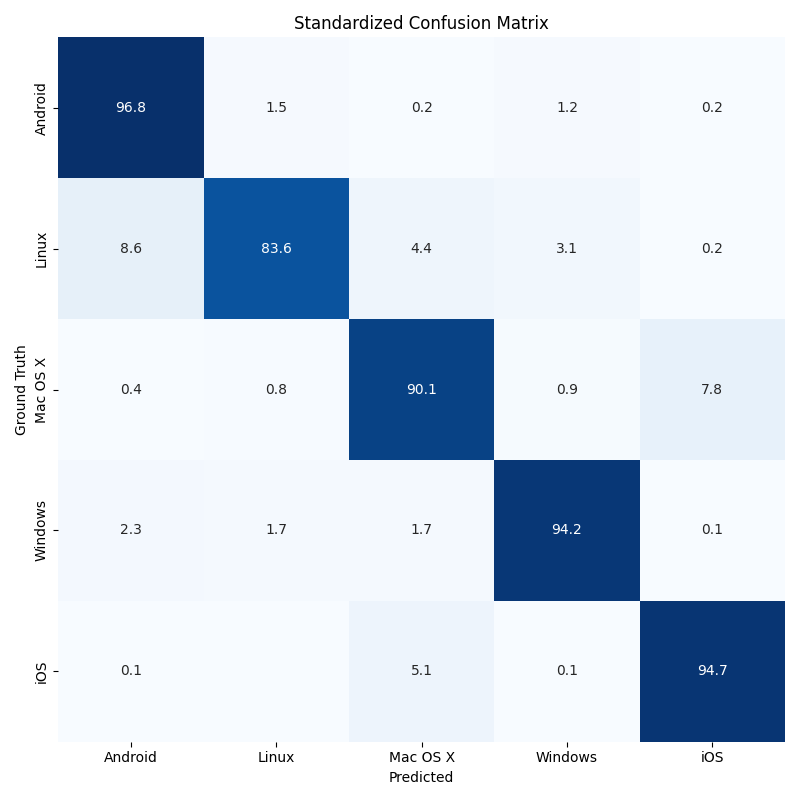}
        \label{fig:conf-matrix-2a}
    \end{subfigure}
    \begin{subfigure}{0.42\textwidth}
        \centering
        \includegraphics[width=\linewidth]{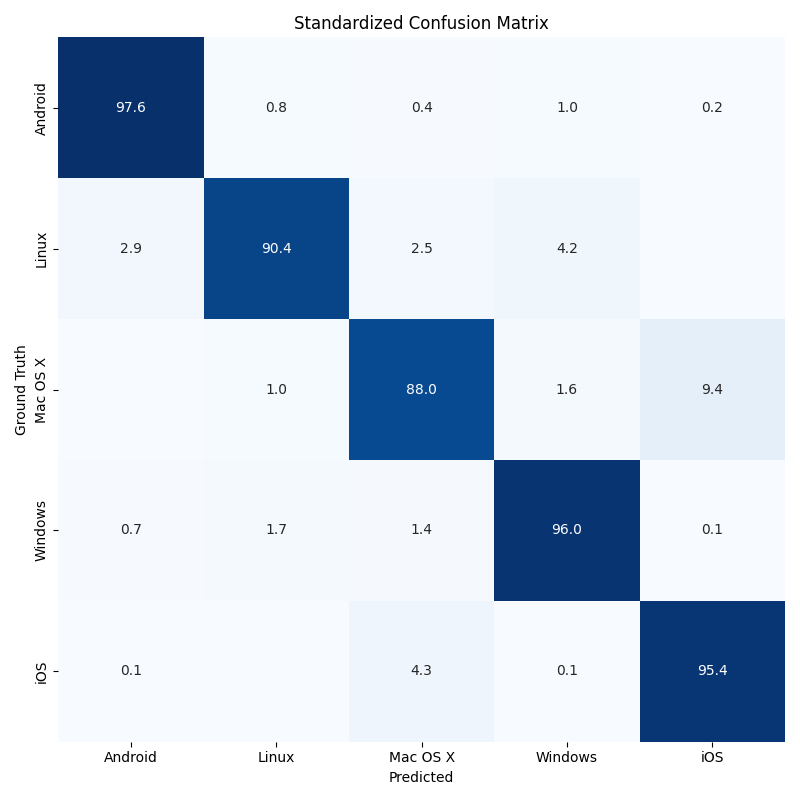}
        \label{fig:conf-matrix-2b}
    \end{subfigure}
    \vspace{-15pt}
    \caption{Standardised confusion matrices for the \texttt{family} classification in \texttt{DAT2} (TabT: \textit{left}, FT-T: \textit{right})}
    \label{fig:conf-matrix-dat2}
\end{figure}

\begin{figure}[]
    \centering
    \begin{subfigure}{0.42\textwidth}
        \centering
        \includegraphics[width=\linewidth]{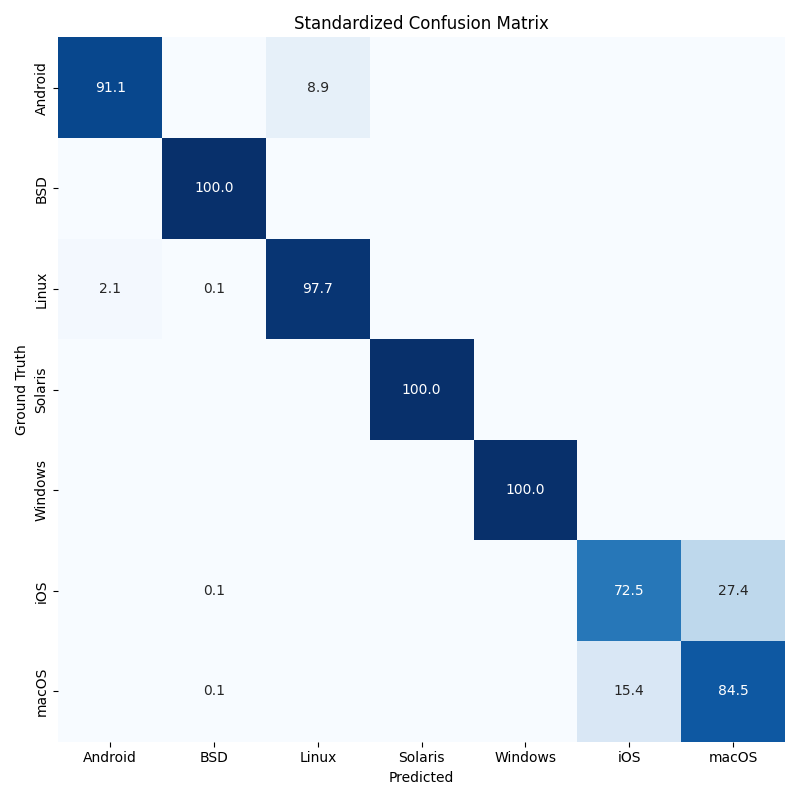}
        \label{fig:conf-matrix-3a}
    \end{subfigure}
    \begin{subfigure}{0.42\textwidth}
        \centering
        \includegraphics[width=\linewidth]{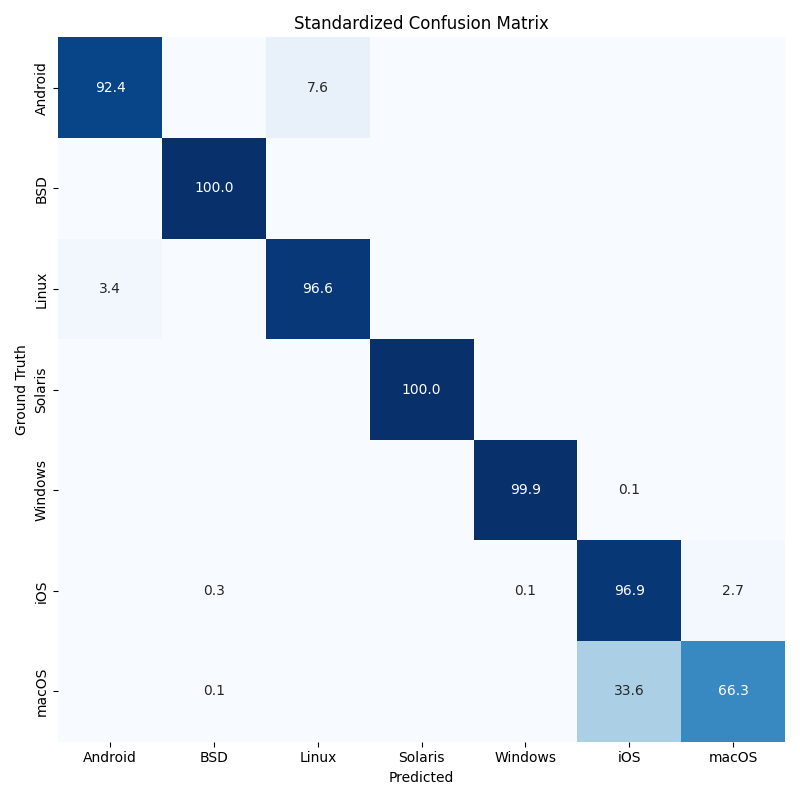}
        \label{fig:conf-matrix-3b}
    \end{subfigure}
    \vspace{-15pt}
    \caption{Standardised confusion matrices for the \texttt{family} classification in \texttt{DAT3} (TabT: \textit{left}, FT-T: \textit{right})}
    \label{fig:conf-matrix-dat3}
\end{figure}

Overall, FT-Transformer emerged as the most robust model, consistently outperforming both TabTransformer and traditional ML methods across most scenarios. A visual comparison of the results is provided in \Cref{fig:results}, and standardized confusion matrices for the \texttt{family} classification are presented in Figures \ref{fig:conf-matrix-dat1}-\ref{fig:conf-matrix-dat3}.

\paragraph{Novel Contributions.} This study pioneers the application of the attention mechanism via Transformer architectures to the OS fingerprinting task, achieving improved outcomes over classic ML models and opening new research directions in advanced deep learning applications for network security.

\paragraph{Reproducibility.} The complete code and all results are publicly available under a GNU GPL v3.0 license at \url{https://github.com/rubenpjove/tabularT-OS-fingerprinting}.

\paragraph{Key Findings and Observations.} Our experimental evaluation on the three datasets demonstrates that Transformer-based models—especially the FT-Transformer (FT-T)—consistently deliver superior performance in OS fingerprinting tasks.

Comparing original and reproduced results in OS fingerprinting research is challenging due to the lack of accuracy metrics or detailed evaluation data, combination of similar OS versions in the same class, etc. This can artificially inflate accuracy rates, leading to inconsistent identification, where some OS versions are specified in detail while others are grouped broadly. Additionally, this study compares the results of the proposed Tabular Transformer-based methods with prior research in the field.

\clearpage

\vfill

\begin{table}[]
\centering
\caption{Best hyperparameters combination for each type of Tabular Transformer, classification task and dataset}
\vspace{5pt}
\resizebox{\textwidth}{!}{
\begin{tabular}{cccccccccc}
\hline
\multirow{2}{*}{\textbf{Dataset}} & \multirow{2}{*}{\textbf{Experiment}} & \multirow{2}{*}{\textbf{Architecture}} & \multicolumn{7}{c}{\textbf{Hyperparameter}}                                                                                                                                                                             \\
                                  &                                      &                                        & \texttt{l\_rate} & \texttt{e\_dim} & \texttt{depth} & \texttt{heads} & \texttt{attn\_d} & \texttt{ff\_d} & \texttt{shared\_categ} \\ \hline
\multirow{2}{*}{\texttt{DAT1}}             & \multirow{2}{*}{\texttt{family}}              & TabT                                   & 0.00010                 & 16                      & 5              & 8              & 0.05                   & 0.20                 & False                              \\
                                  &                                      & {\cellcolor{my_grey}}FT-T                                   & {\cellcolor{my_grey}}0.00198                 & {\cellcolor{my_grey}}32                      & {\cellcolor{my_grey}}2              & {\cellcolor{my_grey}}4              & {\cellcolor{my_grey}}0.15                   & {\cellcolor{my_grey}}0.25                 & {\cellcolor{my_grey}}True                               \\ \hline
\multirow{6}{*}{\texttt{DAT2}}             & \multirow{2}{*}{\texttt{family}}              & TabT                                   & 0.00072                 & 32                      & 4              & 2              & 0.15                   & 0.50                 & True                               \\
                                  &                                      & {\cellcolor{my_grey}}FT-T                                   & {\cellcolor{my_grey}}0.00105                 & {\cellcolor{my_grey}}32                      & {\cellcolor{my_grey}}6              & {\cellcolor{my_grey}}4              & {\cellcolor{my_grey}}0.35                   & {\cellcolor{my_grey}}0.45                 & {\cellcolor{my_grey}}True                               \\
                                  & \multirow{2}{*}{\texttt{major}}               & TabT                                   & 0.00133                 & 16                      & 4              & 2              & 0.05                   & 0.05                 & True                               \\
                                  &                                      & {\cellcolor{my_grey}}FT-T                                   & {\cellcolor{my_grey}}0.00105                 & {\cellcolor{my_grey}}32                      & {\cellcolor{my_grey}}6              & {\cellcolor{my_grey}}4              & {\cellcolor{my_grey}}0.35                   & {\cellcolor{my_grey}}0.45                 & {\cellcolor{my_grey}}True                               \\
                                  & \multirow{2}{*}{\texttt{minor}}               & TabT                                   & 0.00072                 & 16                      & 2              & 8              & 0.15                   & 0.10                 & False                              \\
                                  &                                      & {\cellcolor{my_grey}}FT-T                                   & {\cellcolor{my_grey}}0.00208                 & {\cellcolor{my_grey}}32                      & {\cellcolor{my_grey}}3              & {\cellcolor{my_grey}}6              & {\cellcolor{my_grey}}0.25                   & {\cellcolor{my_grey}}0.15                 & {\cellcolor{my_grey}}True                               \\ \hline
\multirow{4}{*}{\texttt{DAT3}}             & \multirow{2}{*}{\texttt{family}}              & TabT                                   & 0.00039                 & 16                      & 2              & 6              & 0.05                   & 0.50                 & True                               \\
                                  &                                      & {\cellcolor{my_grey}}FT-T                                   & {\cellcolor{my_grey}}0.00198                 & {\cellcolor{my_grey}}32                      & {\cellcolor{my_grey}}2              & {\cellcolor{my_grey}}4              & {\cellcolor{my_grey}}0.15                   & {\cellcolor{my_grey}}0.25                 & {\cellcolor{my_grey}}True                               \\
                                  & \multirow{2}{*}{\texttt{minor}}               & TabT                                   & 0.00028                 & 16                      & 3              & 4              & 0.45                   & 0.25                 & False                              \\
                                  &                                      & {\cellcolor{my_grey}}FT-T                                   & {\cellcolor{my_grey}}0.00208                 & {\cellcolor{my_grey}}32                      & {\cellcolor{my_grey}}3              & {\cellcolor{my_grey}}6              & {\cellcolor{my_grey}}0.25                   & {\cellcolor{my_grey}}0.15                 & {\cellcolor{my_grey}}True                               \\ \hline
\end{tabular}%
}
\label{tab:best-hyps}
\end{table}

\vfill

\begin{table}[]
\centering
\caption{Metrics results for experiments in \texttt{DAT1}}
\vspace{5pt}
\begin{tabular}{ccccccc} 
\hline\hline
\multirow{2}{*}{}                  & \multirow{2}{*}{\textbf{Model}}   & \multicolumn{5}{c}{\textbf{Metric}}                                                                                                                                                                                                                     \\
                                 & \multicolumn{1}{c}{} & \textbf{Accuracy}              & \textbf{Balanced Acc.}        & \textbf{Precision}            & \textbf{Recall}               & \textbf{F1-score}                 \\ 
\hline\hline
\multirow{5}{*}{\rotatebox{90}{\textbf{\texttt{family}}}} & {\cellcolor{my_grey}}\textbf{kNN}                      & {\cellcolor{my_grey}}59.28\%                        & {\cellcolor{my_grey}}56.17\%                       & {\cellcolor{my_grey}}59.99\%                       & {\cellcolor{my_grey}}59.28\%                        & {\cellcolor{my_grey}}58.11\%                           \\
& \textbf{RF}                       & 88.42\%                        & 78.26\%                       & 87.57\%                       & 88.42\%                        & 87.93\%                           \\
& {\cellcolor{my_grey}}\textbf{MLP}                      & {\cellcolor{my_grey}}89.96\%                        & {\cellcolor{my_grey}}80.61\%                       & {\cellcolor{my_grey}}89.68\%                       & {\cellcolor{my_grey}}89.96\%                        & {\cellcolor{my_grey}}89.45\%                           \\
\hhline{~------}
& \textbf{TabT}                     & 89.20\%                        & 82.87\%                       & 89.96\%                       & 89.20\%                        & 89.46\%                           \\
& {\cellcolor{my_grey}}\textbf{FT-T}                     & {\cellcolor{my_blue}\textbf{90.69\%}} & {\cellcolor{my_blue}\textbf{83.41\%}} & {\cellcolor{my_blue}\textbf{91.01\%}} & {\cellcolor{my_blue}\textbf{90.69\%}} & {\cellcolor{my_blue}\textbf{90.80\%}}  \\ 
\hline\hline
\end{tabular}
\label{tab:metrics-DAT1}
\end{table}

\vfill

\begin{table}[]
\centering
\caption{Metrics results for experiments in \texttt{DAT2}}
\vspace{5pt}
\begin{tabular}{ccccccc} 
\hline\hline
\multirow{2}{*}{}                  & \multirow{2}{*}{\textbf{Model}}   & \multicolumn{5}{c}{\textbf{Metric}}                                                                                                                                                                                                                     \\
                                 & \multicolumn{1}{c}{} & \textbf{Accuracy}              & \textbf{Balanced Acc.}        & \textbf{Precision}            & \textbf{Recall}               & \textbf{F1-score}                 \\ 
\hline\hline
\multirow{5}{*}{\rotatebox{90}{\textbf{\texttt{family}}}} & {\cellcolor{my_grey}}\textbf{kNN}                      & {\cellcolor{my_grey}}91.60\%                        & {\cellcolor{my_grey}}90.28\%                       & {\cellcolor{my_grey}}92.54\%                       & {\cellcolor{my_grey}}91.60\%                        & {\cellcolor{my_grey}}91.89\%                           \\
& \textbf{RF}                       & {\cellcolor{my_blue}}\textbf{95.35\%}                        & 92.45\%                       & {\cellcolor{my_blue}}\textbf{95.52\%}                       & {\cellcolor{my_blue}}\textbf{95.35\%}                        & {\cellcolor{my_blue}}\textbf{95.33\%}                           \\
& {\cellcolor{my_grey}}\textbf{MLP}                      & {\cellcolor{my_grey}}93.99\%                        & {\cellcolor{my_grey}}90.78\%                       & {\cellcolor{my_grey}}94.13\%                       & {\cellcolor{my_grey}}93.99\%                        & {\cellcolor{my_grey}}93.99\%                           \\
\hhline{~------}
& \textbf{TabT}                     & 93.86\%                        & 91.90\%                       & 94.10\%                       & 93.86\%                        & 93.93\%                           \\
& {\cellcolor{my_grey}}\textbf{FT-T}                     & {\cellcolor{my_grey}}95.04\% & {\cellcolor{my_blue}\textbf{93.47\%}} & {\cellcolor{my_grey}}95.19\% & {\cellcolor{my_grey}}95.04\% & {\cellcolor{my_grey}}95.09\%  \\ 
\hline\hline
\multirow{5}{*}{\rotatebox{90}{\textbf{\texttt{major}}}} & {\cellcolor{my_grey}}\textbf{kNN}                      & {\cellcolor{my_grey}}70.41\%                        & {\cellcolor{my_grey}}71.89\%                       & {\cellcolor{my_grey}}75.81\%                       & {\cellcolor{my_grey}}70.41\%                        & {\cellcolor{my_grey}}71.79\%                           \\
& \textbf{RF}                       & 73.28\%                        & 66.93\%                       & {\cellcolor{my_blue}}\textbf{89.86\%}                       & 73.28\%                        & 75.64\%                           \\
& {\cellcolor{my_grey}}\textbf{MLP}                      & {\cellcolor{my_grey}}73.59\%                        & {\cellcolor{my_grey}}67.41\%                       & {\cellcolor{my_grey}}78.50\%                       & {\cellcolor{my_grey}}73.59\%                        & {\cellcolor{my_grey}}75.10\%                           \\
\hhline{~------}
& \textbf{TabT}                     & 74.46\%                        & 72.07\%                       & 76.25\%                       & 74.46\%                        & 74.99\%                           \\
& {\cellcolor{my_grey}}\textbf{FT-T}                     & {\cellcolor{my_blue}\textbf{79.09\%}} & {\cellcolor{my_blue}\textbf{80.28\%}} & {\cellcolor{my_grey}}80.00\% & {\cellcolor{my_blue}\textbf{79.09\%}} & {\cellcolor{my_blue}\textbf{79.32\%}}  \\ 
\hline\hline
\multirow{5}{*}{\rotatebox{90}{\textbf{\texttt{minor}}}} & {\cellcolor{my_grey}}\textbf{kNN}                      & {\cellcolor{my_grey}}58.92\%                        & {\cellcolor{my_grey}}62.73\%                       & {\cellcolor{my_grey}}67.65\%                       & {\cellcolor{my_grey}}58.92\%                        & {\cellcolor{my_grey}}60.63\%                           \\
& \textbf{RF}                       & 61.80\%                        & 57.48\%                       & {\cellcolor{my_blue}\textbf{86.58\%}}                       & 61.80\%                        & 65.70\%                           \\
& {\cellcolor{my_grey}}\textbf{MLP}                      & {\cellcolor{my_grey}}61.52\%                        & {\cellcolor{my_grey}}58.27\%                       & {\cellcolor{my_grey}}70.50\%                       & {\cellcolor{my_grey}}61.52\%                        & {\cellcolor{my_grey}}63.92\%                           \\
\hhline{~------}
& \textbf{TabT}                     & 67.74\%                        & 63.50\%                       & 69.49\%                       & 67.74\%                        & 68.28\%                           \\
& {\cellcolor{my_grey}}\textbf{FT-T}                     & {\cellcolor{my_blue}\textbf{68.52\%}} & {\cellcolor{my_blue}\textbf{69.31\%}} & {\cellcolor{my_grey}72.66\%} & {\cellcolor{my_blue}\textbf{68.52\%}} & {\cellcolor{my_blue}\textbf{69.76\%}}  \\ 
\hline\hline
\end{tabular}
\label{tab:metrics-DAT2}
\end{table}

\vfill

\clearpage

\vfill

\begin{table}[]
\centering
\caption{Metrics results for experiments in \texttt{DAT3}}
\vspace{5pt}
\begin{tabular}{ccccccc} 
\hline\hline
\multirow{2}{*}{}                  & \multirow{2}{*}{\textbf{Model}}   & \multicolumn{5}{c}{\textbf{Metric}}                                                                                                                                                                                                                     \\
                                 & \multicolumn{1}{c}{} & \textbf{Accuracy}              & \textbf{Balanced Acc.}        & \textbf{Precision}            & \textbf{Recall}               & \textbf{F1-score}                 \\ 
\hline\hline
\multirow{5}{*}{\rotatebox{90}{\textbf{\texttt{family}}}} & {\cellcolor{my_grey}}\textbf{kNN}                      & {\cellcolor{my_grey}}91.00\%                        & {\cellcolor{my_grey}}92.31\%                       & {\cellcolor{my_grey}}91.23\%                       & {\cellcolor{my_grey}}91.00\%                        & {\cellcolor{my_grey}}91.05\%                           \\
& \textbf{RF}                       & 91.64\%                        & 92.54\%                       & 92.37\%                       & 91.64\%                        & 91.62\%                           \\
& {\cellcolor{my_grey}}\textbf{MLP}                      & {\cellcolor{my_grey}}91.65\%                        & {\cellcolor{my_grey}}92.70\%                       & {\cellcolor{my_grey}}93.50\%                       & {\cellcolor{my_grey}}91.65\%                        & {\cellcolor{my_grey}}91.62\%                           \\
\hhline{~------}
& \textbf{TabT}                     & 91.45\%                        & 92.27\%                       & 91.64\%                       & 91.45\%                        & 91.43\%                           \\
& {\cellcolor{my_grey}}\textbf{FT-T}                     & {\cellcolor{my_blue}\textbf{92.35\%}} & {\cellcolor{my_blue}\textbf{93.15\%}} & {\cellcolor{my_blue}\textbf{93.74\%}} & {\cellcolor{my_blue}\textbf{92.35\%}} & {\cellcolor{my_blue}\textbf{92.23\%}}  \\ 
\hline\hline
\multirow{5}{*}{\rotatebox{90}{\textbf{\texttt{minor}}}} & {\cellcolor{my_grey}}\textbf{kNN}                      & {\cellcolor{my_grey}}74.42\%                        & {\cellcolor{my_grey}}63.49\%                       & {\cellcolor{my_grey}}76.68\%                       & {\cellcolor{my_grey}}74.42\%                        & {\cellcolor{my_grey}}74.92\%                           \\
& \textbf{RF}                       & 73.92\%                        & 60.19\%                       & {\cellcolor{my_blue}\textbf{81.76\%}} & 73.92\%                        & 75.66\%                           \\
& {\cellcolor{my_grey}}\textbf{MLP}                      & {\cellcolor{my_grey}}72.94\%                        & {\cellcolor{my_grey}}62.19\%                       & {\cellcolor{my_grey}}75.89\%                       & {\cellcolor{my_grey}}72.94\%                        & {\cellcolor{my_grey}}74.14\%                           \\
\hhline{~------}
& \textbf{TabT}                     & {\cellcolor{my_blue}\textbf{75.59\%}}                        & {\cellcolor{my_blue}\textbf{64.11\%}}                       & 81.68\%                       & {\cellcolor{my_blue}\textbf{75.59\%}}                        & {\cellcolor{my_blue}\textbf{75.90\%}}                           \\
& {\cellcolor{my_grey}}\textbf{FT-T}                     & {\cellcolor{my_grey}}75.33\% & {\cellcolor{my_grey}}63.70\% & {\cellcolor{my_grey}}76.92\% & {\cellcolor{my_grey}}75.33\% & {\cellcolor{my_grey}}75.74\%  \\ 
\hline\hline
\end{tabular}
\label{tab:metrics-DAT3}
\end{table}

\vfill

\vfill

\clearpage


\clearpage

\Cref{tab:comparison} offers a comprehensive comparison of the proposed methods against existing techniques, for each employed dataset (\texttt{DAT1}, \texttt{DAT2} and \texttt{DAT3}) and granularity of the classification (\texttt{family}, \texttt{major} and \texttt{minor}). For each work, we specify the ML technique used.

\begin{itemize}
    \item \textbf{\texttt{DAT1}:} \emph{FT-Transformer outperforms the state-of-the-art method by approximately 12\% in F1-score.} This dataset, originally employed in \cite{lastovicka_usingTLS_2020}, contains only OS information for \texttt{family} classification. While the referenced work compared a novel TLS-based decision tree with select baselines, our FT-T method achieves an F1-score improvement of around 12\%, as shown in the first part of \Cref{tab:comparison}.
    
    \item \textbf{\texttt{DAT2}:} \emph{FT-Transformer sets a new benchmark for \texttt{family} classification, uniquely addresses the \texttt{major} level, and remains competitive on the challenging \texttt{minor} task.} Utilized in \cite{lastovicka_passive_2023}, \texttt{DAT2} supports OS classification at \texttt{family}, \texttt{major}, and \texttt{minor} levels. For \texttt{family} classification, FT-T achieves an F1-score of 95.09\%, substantially outperforming the second-best model (DT at 81.60\%). No prior work has addressed the \texttt{major} level, making FT-T the sole and best-performing model in this category. For the \texttt{minor} classification, despite its inherent difficulty, FT-T remains competitive with an F1-score of 69.76\%.
    
    \item \textbf{\texttt{DAT3}:} \emph{FT-Transformer excels in \texttt{family} classification while TabTransformer shows the best performance on \texttt{minor} classification.} Derived from the Nmap (v.7.94) database, \texttt{DAT3} supports both \texttt{family} and \texttt{minor} classifications. In the \texttt{family} task, FT-T achieves an F1-score of 92.23\%, outperforming the best baseline (MLP). Conversely, for the \texttt{minor} classification, TabTransformer outperforms the kNN baseline with an F1-score of 75.90\%.
\end{itemize}

Overall, the FT-Transformer's consistent superior performance suggests it is particularly adept at extracting informative features from both numerical and categorical data through its attention mechanism.

\vfill

\begin{table}[h!]
\centering
\caption{Comparative analysis of proposed methods and previous works for OS fingerprinting across various datasets and classification tasks.}
\vspace{5pt}
\label{tab:comparison}
\begin{tabular}{cccccccc}
\hline \hline
\textbf{Dataset}       & \textbf{Classification}          & \textbf{Source} & \textbf{Model} & \textbf{Accuracy} & \textbf{Precision} & \textbf{Recall} & \textbf{F1-Score} \\ \hline \hline
\multirow{2}{*}{\texttt{DAT1}}  & \multirow{2}{*}{\texttt{family}} & {\cellcolor{my_grey}}\cite{lastovicka_usingTLS_2020} & {\cellcolor{my_grey}}DT             & {\cellcolor{my_blue}}\textbf{93.12\%}            & {\cellcolor{my_grey}}80.48\%             & {\cellcolor{my_grey}}77.49\%          & {\cellcolor{my_grey}}78.96\%            \\
\hhline{~~------}
                       &                         & Proposed        & FT-T           & 90.69\%            & {\cellcolor{my_blue}}\textbf{91.01\%}             & {\cellcolor{my_blue}}\textbf{90.69\%}          & {\cellcolor{my_blue}}\textbf{90.80\%}            \\ \hline \hline
\multirow{12}{*}{\texttt{DAT2}} & \multirow{6}{*}{\texttt{family}} & {\cellcolor{my_grey}}\cite{beverly_robust_2004} in \cite{lastovicka_passive_2023}        & {\cellcolor{my_grey}}Bayes          & {\cellcolor{my_grey}}37.70\%           & {\cellcolor{my_grey}}49.70\%            & {\cellcolor{my_grey}}37.60\%         & {\cellcolor{my_grey}}37.40\%           \\
                       &                         & \cite{lastovicka_cybersecurity_2018} in \cite{lastovicka_passive_2023}   & DT             & 84.10\%           & 83.60\%            & 94.30\%         & 81.60\%           \\
                       &                         & {\cellcolor{my_grey}}\cite{lastovicka_usingTLS_2020} in \cite{lastovicka_passive_2023}   & {\cellcolor{my_grey}}DT             & {\cellcolor{my_grey}}82.10\%           & {\cellcolor{my_grey}}81.60\%            & {\cellcolor{my_grey}}92.40\%         & {\cellcolor{my_grey}}81.60\%           \\
                       &                         & \cite{lippmann_passive_2003} in \cite{lastovicka_passive_2023}       & kNN            & 94.70\%           & -                  & -               & -                 \\
                       &                         & {\cellcolor{my_grey}}\cite{hulak_evaluation_2023}           & {\cellcolor{my_grey}}DT/RF          & {\cellcolor{my_grey}}88.00\%           & {\cellcolor{my_grey}}76.00\%            & {\cellcolor{my_grey}}-               & {\cellcolor{my_grey}}76.00\%           \\
\hhline{~~------}
                       &                         & Proposed        & FT-T           & {\cellcolor{my_blue}}\textbf{95.04\%}           & {\cellcolor{my_blue}}\textbf{95.19\%}            & {\cellcolor{my_blue}}\textbf{95.04\%}         & {\cellcolor{my_blue}}\textbf{95.09\%}           \\ 
                       \hhline{~-------}
                       & \texttt{major}                   & {\cellcolor{my_grey}}Proposed        & {\cellcolor{my_grey}}FT-T           & {\cellcolor{my_blue}}\textbf{79.09\%}           & {\cellcolor{my_blue}}\textbf{80.00\%}            & {\cellcolor{my_blue}}\textbf{79.09\%}         & {\cellcolor{my_blue}}\textbf{79.32\%}           \\ 
                       \hhline{~-------}
                       & \multirow{5}{*}{\texttt{minor}}  & \cite{beverly_robust_2004} in \cite{lastovicka_passive_2023}         & Bayes          & 0.50\%            & {\cellcolor{my_blue}}\textbf{92.10\%}            & 0.60\%          & 0.30\%            \\ 
                       &                         & {\cellcolor{my_grey}}\cite{lastovicka_cybersecurity_2018} in \cite{lastovicka_passive_2023}   & {\cellcolor{my_grey}}DT             & {\cellcolor{my_grey}}73.70\%           & {\cellcolor{my_grey}}81.10\%            & {\cellcolor{my_blue}}\textbf{73.60\%}         & {\cellcolor{my_grey}}68.50\%           \\
                       &                         & \cite{lastovicka_usingTLS_2020} in \cite{lastovicka_passive_2023}   & DT             & 73.40\%           & 71.50\%            & 73.40\%         & {\cellcolor{my_blue}}\textbf{71.40\%}           \\
                       &                         & {\cellcolor{my_grey}}\cite{lippmann_passive_2003} in \cite{lastovicka_passive_2023}       & {\cellcolor{my_grey}}kNN            & {\cellcolor{my_blue}}\textbf{92.10\%}           & {\cellcolor{my_grey}}-                  & {\cellcolor{my_grey}}-               & {\cellcolor{my_grey}}-                 \\
\hhline{~~------}
                       &                         & Proposed        & FT-T           & 68.52\%           & 72.66\%            & 68.52\%         & 69.76\%           \\ \hline \hline
\multirow{4}{*}{\texttt{DAT3}}  & \multirow{2}{*}{\texttt{family}} & {\cellcolor{my_grey}}Baseline        & {\cellcolor{my_grey}}MLP            & {\cellcolor{my_grey}}91.65\%           & {\cellcolor{my_grey}}93.50\%            & {\cellcolor{my_grey}}91.65\%         & {\cellcolor{my_grey}}91.62\%           \\
\hhline{~~------}
                       &                         & Proposed        & FT-T           & {\cellcolor{my_blue}}\textbf{92.35\%}           & {\cellcolor{my_blue}}\textbf{93.74\%}            & {\cellcolor{my_blue}}\textbf{92.35\%}         & {\cellcolor{my_blue}}\textbf{92.23\%}           \\ 
                       \hhline{~-------}
                       & \multirow{2}{*}{\texttt{minor}}  & {\cellcolor{my_grey}}Baseline        & {\cellcolor{my_grey}}RF            & {\cellcolor{my_grey}}73.92\%           & {\cellcolor{my_blue}}\textbf{81.76\%}            & {\cellcolor{my_grey}}73.92\%         & {\cellcolor{my_grey}}75.66\%           \\
\hhline{~~------}
                       &                         & Proposed        & Tab-T          & {\cellcolor{my_blue}}\textbf{75.59\%}           & 81.68\%            & {\cellcolor{my_blue}}\textbf{75.59\%}         & {\cellcolor{my_blue}}\textbf{75.90\%}           \\ 
\hline \hline
\end{tabular}
\end{table}

\vfill

\clearpage

%% file: content/s06-conclusion.tex
\section{Conclusion \& Future Work}
\label{sec:conclusion_future_work}

In this work, we introduced the novel application of Transformer architectures adapted for tabular data to the task of OS fingerprinting. By leveraging both the TabTransformer and FT-Transformer models, our study bridges the gap between traditional rule-based methods and modern DL approaches in the cybersecurity domain. The experimental evaluation across multiple diverse datasets demonstrated that attention-based models can effectively capture complex interactions within network traffic, yielding improved OS classification performance over conventional ML baselines and previous works.

Our results indicate that the FT-Transformer, in particular, offers significant advantages in terms of accuracy and robustness when classifying OSs at varying levels of granularity and network environments. The inherent self-attention mechanisms allow the model to learn intricate feature dependencies, which is crucial for processing heterogeneous and dynamic network traffic data. This performance gain, as compared to previous works and traditional models such as k-Nearest Neighbours, Random Forests, and Multi-Layer Perceptron, establishes a new benchmark for OS fingerprinting tasks and highlights the potential of advanced DL techniques in this area.

In future work, we aim to explore several directions to further enhance the application of Tabular Transformer architectures for OS fingerprinting. First, expanding the range of datasets. Although the selected datasets are, in principle, representative of modern network environments, additional data from real-world deployments would be beneficial to further validate the models under a wider range of conditions. Additionally, we plan to investigate the impact of incorporating hybrid DL approaches, such as combining Transformers with Graph Neural Networks to better capture the temporal–spatial relationships in network traffic data. Furthermore, we intend to evaluate the performance of these architectures on new or modified OS versions, directly comparing them with traditional tools such as Nmap, to assess their robustness and practical applicability in dynamic environments. Finally, we aim to apply transfer learning by pre-training on larger, related datasets to further refine performance across diverse and dynamic network conditions.

Moreover, the integration of the developed model into real-world cybersecurity tools, such as Nmap and various intrusion detection systems, could substantially enhance their versatility and operational effectiveness. We believe that our approach is not only beneficial for OS fingerprinting but also holds promise for broader applications within network security, including intrusion detection and traffic anomaly detection, thereby potentially yielding a significant industrial impact.